\documentclass[aps,prd,showpacs,nofootinbib,superscriptaddress,amsmath,amssymb]{revtex4}

\usepackage{array} 
\usepackage{braket}

\usepackage{tikz}
\usetikzlibrary{fadings}
\usetikzlibrary{patterns}
\usetikzlibrary{shadows.blur}
\usetikzlibrary{shapes}

\usepackage{definitions}%

\begin{document}

\title{Chiral Separation effect in non - homogeneous systems}

\author{M. Suleymanov}
\affiliation{Physics Department, Ariel University, Ariel 40700, Israel}

\author{M. A. Zubkov\footnote{On leave of absence from NRC "Kurchatov Institute" - ITEP, B. Cheremushkinskaya 25, Moscow, 117259, Russia}}
\affiliation{Physics Department, Ariel University, Ariel 40700, Israel}

\begin{abstract}
We discuss chiral separation effect in the systems with spatial  non - homogeneity. It may be caused by non - uniform electric potential or by another reasons, which do not, however, break chiral symmetry of an effective low energy theory. Such low energy effective theory describes quasiparticles close to the Fermi surfaces. In the presence of constant external magnetic field the non - dissipative axial current appears. It appears that its response to chemical potential and magnetic field (the CSE conductivity) is universal. It is robust to smooth modifications of the system and is expressed through an integral over a surface in momentum space that surrounds all singularities of the Green function. In itself this expression represents an extension of the topological invariant protecting Fermi points to the case of inhomogeneous systems.
\end{abstract}



\maketitle
\tableofcontents

\section{Introduction}

In the recent years the non - dissipative transport effects attract attention both in the framework of condensed matter physics and in the high energy physics \cite{Landsteiner:2012kd,semimetal_effects7,Gorbar:2015wya,Miransky:2015ava,Valgushev:2015pjn,Buividovich:2015ara,Buividovich:2014dha,Buividovich:2013hza}. An important arena for the experimental observation of these effects is given by Dirac and Weyl semimetals \cite{semimetal_effects6,semimetal_effects10,semimetal_effects11,semimetal_effects12,semimetal_effects13,Zyuzin:2012tv,tewary,16}.
These materials also represent the bridge between  high energy theory and condensed matter physics because the physics of fermionic quasiparticles inside them models physics of elementary particles.
The chiral separation effect (CSE) is one of the non - dissipative transport effects. It has been proposed by M.Metlitski and A.Zhitnitsky \cite{Metl}. The essence of this effect is appearance of a non - dissipative axial current in the direction of external magnetic field. The original calculations of this effect have been performed in the system of continuum Dirac fermions. Without external magnetic field this system is homogeneous. It has been found that in the chiral limit (when Dirac fermions are massless) the axial current is proportional to the external magnetic field strength $F_{ij}$ and the ordinary chemical potential $\mu$ counted from the Fermi point (the point in momentum space where fermion energy levels cross each other):
\begin{equation}
J_5^k = -\frac{1}{4\pi^2}\epsilon^{ijk0} \mu F_{ij}\label{1}
\end{equation}
The theoretical prediction of this effect has been followed by the number of papers discussing the possibility to observe it during heavy ion collisions \cite{Kharzeev:2015znc,Kharzeev:2009mf,Kharzeev:2013ffa,ref:HIC}. The fireballs appeared during the heavy ion collisions are widely believed to contain the new state of matter called quark  - gluon plasma  \cite{QCDphases,1,2,3,4,5,6,7,8,9,10}. In this state quarks are free and almost massless. Their masses may actually be neglected completely and we may speak of the system of true chiral fermions. External magnetic field appears here because the ion beams carry electric current. During the non - central collisions the two colliding ions, therefore, produce strong magnetic field orthogonal to their trajectories. The CSE results in the appearance of axial current within the fireball. After decay of the fireball the asymmetry in the distribution of outgoing particles carries the signature of the chiral separation effect. It is worth mentioning, that the CSE as well as its cousin - the chiral vortical effect (CVE) may also be relevant for the description of the quark matter under extreme conditions in the rotated neutron stars~\cite{Cook:1993qr}.

Actually, the CSE represents a certain incarnation of chiral anomaly. This relation has been discussed in a number of papers (see, for example, \cite{Zyuzin:2012tv}). The similar conjecture has been proposed for the so - called Chiral Magnetic Effect  (CME) \cite{Vilenkin,CME,Kharzeev:2013ffa,Kharzeev:2009pj,SonYamamoto2012}. The important difference between the two is, however, that the true equilibrium theory does not admit the presence of the CME \cite{Valgushev:2015pjn,Buividovich:2015ara,Buividovich:2014dha,Buividovich:2013hza,Z2016_1,Z2016_2,nogo,nogo2} while the CSE exists as an equilibrium phenomenon \cite{KZ2017}. Out of equilibrium, however, the CME is back \cite{Nielsen:1983rb}, which, possibly, manifests itself in experiments with Dirac semimetals \cite{ZrTe5}.

We would like to notice several recent works on the CSE. In the framework of continuum  quantum field theory the CSE was discussed, for example, in \cite{Gorbar:2015wya}. The lattice regularization has been used in \cite{Buividovich:2013hza} and in \cite{KZ2017}. It has been shown that if the model is considered at small but finite temperatures, then the lattice regularization gives the conventional result for the CSE. It appears that in the framework of continuum theory the order of summation over Matsubara frequencies and integration over momenta is important. The uncertainty appears if integration over the $3$ - momenta is performed first. This explains the importance of lattice regularization  for the investigation of the CSE. \cite{KZ2017} reports conventional expression for the conductivity of CSE in the lattice models, which describe chiral fermions at low energies.  In the presence of the finite mass of the fermions the expression for the CSE is changed. On the formal level the theory with massless fermions suffers from various infrared divergencies \cite{infrared1,infrared2,infrared3}. For this reason  the finite fermion masses are worth to be introduced even if the limit of small masses is assumed. Notice, that in \cite{Vilenkin} the neutral particles were discussed, and for them there are no infrared divergencies related to radiation of photons. Interaction corrections to CSE have been considered, for example, in  \cite{Shovkovy}. It has been argued that the higher orders of perturbation theory do give corrections to the version of the CSE of massive fermions.

To the best of our knowledge the CSE has not been considered in sufficient details for the inhomogeneous systems. At the same time in any real situation, when the CSE takes place, it exists in essentially inhomogeneous systems. In case of the heavy ion collisions the chiral quarks in the fireballs exist in the presence of non - uniform environment. The effective action of the chiral quarks here cannot have the form of a homogeneous Dirac action. Instead it should depend on various background fields that depend on coordinates. The external magnetic field in this problem as well as the quark chemical potential is also coordinate dependent. In case of the CSE in Dirac/Weyl semimetals the more or less uniform external magnetic field can be provided as well as the uniform chemical potential. However, the effective action for electron quasi - particles in itself is always not homogeneous even in the absence of external magnetic field. There are always impurities, and various sources of elastic deformations - both internal (dislocations and disclinations) and external (mechanical stress caused by external forces). In the present paper we concentrate on the latter situation -- uniform external magnetic field and uniform chemical potential, but non - uniform fermionic action. In order to deal with the non - homogeneous systems we use the Weyl - Wigner formalism.

Weyl-Wigner formalism \cite{Weyl,Wigner} as an alternative formulation of non-relativistic quantum mechanics has been developed by H. Groenewold \cite{Groenewold} and J. Moyal \cite{Moyal}.  Later it was  adopted in some form both for the quantum field theory and for the condensed matter physics. It is based on the notions of the Weyl symbol of operator and the Wigner distribution function. The Wigner  - Weyl formalism in quantum mechanics is often referred to as the phase space formulation. It is defined in phase space that is composed of both coordinates and momenta, while, the conventional formulation uses either coordinate space or  momentum space representations. In the phase space formulation the quantum state is described by the Wigner distribution (instead of a wave function), while the product of operators is replaced by the Moyal product of functions defined in phase space.

Lattice field theories were proposed as a mathematical tool to deal with divergences in quantum field theories calculations in high energy physics. On the other hand, in addition to the "traditional", approach of quantum mechanics in solid state physics \cite {Fradkin}, there is a lot of activity of using quantum field theory ideas in condensed matter systems. The attempts to apply the numerical lattice QFT methods in condensed matter started with Monte Carlo simulations of graphene \cite{Drut_Lahde_2009}.

Although there were attempts to construct an exact phase space formalism for lattice theories and finite state quantum systems (Schwinger \cite{Schwinger}, Buot \cite{Buot1,Buot2,Buot3}, Wootters \cite{Wootters}, Leonhardt \cite{Leonhardt}, Kasperkovitz \cite{Kasperkovitz} and Ligabo\cite{Ligabo}), until recently such an approach has not been proposed. It has been developed recently  in \cite{FZ2019_2}. However, the present paper is based on the simplified version of lattice Wigner - Weyl calculus valid for the case when an inhomogeneity is sufficiently weak. This method  is an approximation in case of condensed matter physics, where the lattice describes a real material, but is exact in case of high energy physics where the lattice is a mathematical tool. In the case of condensed matter systems, it is shown that this approximation holds for any physically reasonable fields from the experimental point of view. In this approach, Weyl-Wigner phase space formalism is used to calculate Dirac operators and Green's functions. These techniques are widely used in recent research \cite{Zhang_Zubkov_PRD_2019,Suleymanov_Zubkov_2019,Fialk_Zubkov_2020_sym,Zhang_Zubkov_JETP_2019,Zhang_Zubkov_PhysLet_2020} dealing with linear response to electromagnetic fields which are shown to be topological invariants, as quantum Hall conductance for example.

We consider the lattice tight - binding models of a rather general type. In these models the fermions are placed in four - component Dirac spinors. Extra internal indices of these spinors are also admitted (valley indices, real spin indices etc). As a result action for the fermionic quasiparticles contains the structure of $4\times 4$ matrix to be expressed through Dirac matrices $\gamma^k$ for $k=1,2,3,4,5$, and their derivatives $\sigma^{kj} = \frac{1}{4i}[\gamma^k,\gamma^j]$. We are interested in the situation, when low energy effective theory of such models obeys chiral symmetry. For the homogeneous models this would mean that matrix $\gamma^5$ commutes or anti - commutes with the one - particle Hamiltonian in a small vicinity of Fermi surfaces/Fermi points, where the low energy effective theory arises. The Fermi surface manifests itself in the two - point Green function $\hat{G} = \hat{Q}^{-1}$ as the position of its singularities in momentum space. Here $\hat{Q}$ is the so - called lattice Dirac operator.  For the non - homogeneous systems with sufficiently weak inhomogeneity operator $\hat{Q}$ is not diagonal in momentum, and the notion of ordinary Fermi surface/Fermi point may be replaced by the coordinate - dependent Fermi surface/Fermi point \cite{Volovik2003}. The space dependent Fermi point is known also as emergent gauge field. Mathematically the weakness of inhomogeneity means that the poles of the Wigner transformed two - point Green function $G_W(p,x)$ at each value of $x$ are given by zeros of the Weyl symbol $Q_W(p,x)$ of operator $\hat{Q}$. (The precise definitions of Wigner transformation and Weyl symbol will be given in the next Section of the present paper.)  For the case when inhomogeneity is more strong the zeros of $Q_W$ do not coincide with the poles of $G_W$. An extension of the notion of Fermi surface to this case may be given by the hyper - surface in phase space (space of both coordinates $x$ and momenta $p$). We may choose the hyper - surface, where $Q_W(p,x) = 0$. The other possible definition is localization of the singularities of $G_W(p,x)$. One may also consider position of the singularities of a certain combination of $Q_W$ and $G_W$ entering expression for the CSE conductivity (to be specified below), and this is our way for the definition of hypersurface $\Xi$ in phase space, which is the extension of the notion of the Fermi surface. For the non - homogeneous systems the low energy physics appears in a certain vicinity of $\Xi$, and we require that $Q_W$ and $G_W$ commute or anti - commute with $\gamma^5$ in this vicinity. Recall that the precise chiral symmetry cannot be maintained within the whole phase space for the lattice tight - binding models except for the marginal cases. An example of the marginal case is the system of naive lattice Dirac fermions. In this case the $16$ fermion doublers appear, and the sum of their contributions to the CSE conductivity vanishes \cite{KZ2017}.

We show that under the above conditions (chiral symmetry of low energy effective theory) the axial current of CSE in the non - homogeneous system of general type is still proportional to external magnetic field. Being averaged over the whole volume of the system it may be expressed as
\begin{equation}
\bar{J}_5^k = -\frac{\mathcal N}{4\pi^2}\epsilon^{ijk0} \mu F_{ij}\label{1_}
\end{equation}
 where $\mathcal{N}$ is a topological invariant expressed through $G_W$ and $Q_W$. (Here $\mu$ is counted from the level, where $\bar{J}_5=0$.) This expression contains an integral over a surface $\Sigma_3$ in momentum space surrounding the singularities of an expression standing inside the integral (that is the hyper - surface $\Xi$ mentioned above). Being the topological invariant $\mathcal{N}$ is robust to an arbitrary smooth modification of the given system as long as this modification does not break chiral symmetry of quasiparticles existing close to $\Xi$.

All inhomogeneous lattice systems with chiral fermions (those with the chiral symmetry at low energies) may be subdivided into the homotopic classes. Within each class the operators $\hat{Q}$ are connected to each other by smooth deformation. The values of $\mathcal{N}$ are constant within each homotopic class. For the homogeneous representative of each homotopic class the value of $\mathcal{N}$ may be calculated easily. It is given by the number of the species of chiral Dirac fermions in the corresponding low energy effective theory. This allows to calculate easily the CSE conductivity $\mathcal{N}/(2\pi^2)$ for any inhomogeneous system.

In the present paper we do not consider interactions between the quasiparticles. It is worth mentioning that in the case of Integer Quantum Hall effect the similar problem has been considered recently (see \cite{Zhang_Zubkov_JETP_2019}).
It has been shown that the expression for Hall conductivity through $\hat{G}$ has the same form as for the non - interacting case but with $\hat{G}$ replaced by the complete interacting Green function.
Based on the approach of \cite{Zhang_Zubkov_JETP_2019} we expect that the same refers to expression for $\mathcal{N}$ of Eq. (\ref{1_}). However, the consideration of this issue remains out of the scope of the present paper.

\section{Weyl-Wigner phase space formalism}

In this section we briefly review  the technique of  Wigner transformation applied to quantum mechanics  defined in infinite continuous coordinate space.
Phase space formalism allows to describe quantum mechanics using c-functions instead of operators. Weyl-Wigner transformation of the matrix elements of operator ({\it that is Weyl symbol of operator}) represents such a correspondence.

\subsection{Weyl symbol of operator and Wigner distribution function}

We start from definition of an average of  operator $\hat{A}$  with respect to quantum state $\Psi$
\begin{equation}\begin{aligned}
\bra{\Psi} \hat{A} \ket{\Psi}=
&\int\limits_{-\infty}^\infty dx \int\limits_{-\infty}^\infty dy \braket{\Psi|x}  \bra{x} \hat{A} \ket{y}  \braket{y|\Psi}=
\int\limits_{-\infty}^\infty dp \int\limits_{-\infty}^\infty dq \braket{\Psi|p}  \bra{p} \hat{A} \ket{q}  \braket{q|\Psi}
\label{<A>} \end{aligned}\end{equation}
Here by $x, y$ or $p, q$ we denote the continuous coordinates or momentum respectively. For simplicity we consider the case of one - dimensional space $R^1$. The generalization of our expressions to the case of $D$ - dimensional space $R^D$  is straightforward.\\
Let us change the coordinates:
\begin{equation}\begin{aligned}
x=u+v/2 \,\,\,\,\,\,\,\,\, y=u-v/2
\label{Z} \end{aligned} \end{equation}
Then
\begin{equation}\begin{aligned}
dxdy=\frac{\partial(x,y)}{\partial(u,v)}dudv=
\left |
\begin{matrix}
\frac {\partial x}{\partial u} & \frac {\partial y}{\partial u}\\
\frac {\partial x}{\partial v} & \frac {\partial y}{\partial v}
\end{matrix}
\right |
dudv=-dudv
\label{Z} \end{aligned} \end{equation}
This gives
\begin{equation} \begin{aligned}
&\bra{\Psi} \hat{A} \ket{\Psi}
=-\int dx dy \bra{x+y/2} \hat{A} \ket{x-y/2} \braket{\Psi|x+y/2} \braket{x-y/2|\Psi}=\\
-&\int dx dy dz \bra{x+y/2} \hat{A} \ket{x-y/2} \delta(z-y) \braket{\Psi|x+z/2} \braket{x-z/2|\Psi}=\\
-&\int dx dy dz dp \bra{x+y/2} \hat{A} \ket{x-y/2} \frac{e^{ip(z-y)}}{2\pi} \braket{x-z/2|\Psi} \braket{\Psi|x+z/2}= \\
&\int \frac{dx dp}{2\pi} dy e^{-ipy} \bra{x+y/2} \hat{A} \ket{x-y/2} dz e^{ipz} \braket{x+z/2|\Psi} \braket{\Psi|x-z/2}
\label{Z}\end{aligned} \end{equation}
\begin{equation}\begin{aligned}
\bra{\Psi} \hat{A} \ket{\Psi}=\frac{1}{2\pi} \int\limits_{-\infty}^\infty dx \int\limits_{-\infty}^\infty dp A_W(x,p) \rho_W(x,p)
\label{<A>} \end{aligned}\end{equation}

Weyl symbol of operator $A_W(x,p)$ and Wigner distribution $W(x,p)$ are defined as follows (where the momentum space representation may be obtained using the similar way)
\begin{equation} \begin{aligned}
&A_W(x,p)\equiv
\int\limits_{-\infty}^\infty  dy e^{-ipy} \bra{x+\frac{y}{2}} \hat{A} \ket{x-\frac{y}{2}}=
\int\limits_{-\infty}^\infty  dq e^{iqx} \bra{p+\frac{q}{2}} \hat{A} \ket{p-\frac{q}{2}}
\label{A_W}\end{aligned} \end{equation}
\begin{equation} \begin{aligned}
W(x,p)=\int\limits_{-\infty}^\infty dy  e^{ipy} \braket{x-\frac{y}{2}|\Psi} \braket{\Psi|x+\frac{y}{2}}
=\int\limits_{-\infty}^\infty dq  e^{-iqx} \braket{p-\frac{q}{2}|\Psi} \braket{\Psi|p+\frac{q}{2}}
\label{W_W}\end{aligned} \end{equation}

\subsection{Moyal product}

The Weyl symbol of the product of two operators, called the Moyal product, is defined as follows (this time in the momentum representation)
\begin{equation} \begin{aligned}
&(\hat A \hat B)_W=
\int\limits_{-\infty}^\infty  dq e^{iqx} \bra{p+\frac{q}{2}} \hat{A} \hat B \ket{p-\frac{q}{2}}=
\int\limits_{-\infty}^\infty  dk \int\limits_{-\infty}^\infty  dq
e^{iqx} \bra{p+\frac{q}{2}}  \hat{A}
\ket{k}\bra{k} \hat B \ket{p-\frac{q}{2}}
\label{Z}\end{aligned} \end{equation}
changing variables
\begin{equation}\begin{aligned}
q=u+v \,\,\,\,\,\,\,\,\,\,\,\, k=p-u/2+v/2
\label{Z}\end{aligned}\end{equation}
\begin{equation}\begin{aligned}
(\hat A \hat B)_W=&
\int\limits_{-\infty}^\infty  du \int\limits_{-\infty}^\infty  dv
e^{iux} \bra{p+\frac{u}{2}+\frac{v}{2}}  \hat{A} \ket{p-\frac{u}{2}+\frac{v}{2}}
e^{ivx} \bra{p-\frac{u}{2}+\frac{v}{2}} \hat B \ket{p-\frac{u}{2}-\frac{v}{2}}=\\
&\int\limits_{-\infty}^\infty  du \int\limits_{-\infty}^\infty  dv
\left[e^{iux} \bra{p+\frac{u}{2}} \hat{A} \ket{p-\frac{u}{2}}\right]
e^{\frac{i}{2} \left( \overleftarrow{\partial_x}\overrightarrow{\partial_p}-\overleftarrow{\partial_p}\overrightarrow{\partial_x}\right )}
\left[e^{ivx} \bra{p+\frac{v}{2}} \hat B \ket{p-\frac{v}{2}}\right]
\label{Z}\end{aligned} \end{equation}
\begin{equation}\begin{aligned}
(AB)_W(x,p)\equiv A_W(x,p)\star B_W(x,p)=A_W(x,p)e^{\overleftrightarrow{\Delta}} B_W(x,p)
\label{Moyal}\end{aligned}\end{equation}
where
\begin{equation}\begin{aligned}
\overleftrightarrow{\Delta} \equiv{\frac{i}{2} \left( \overleftarrow{\partial}_{x}\overrightarrow{\partial_p}-\overleftarrow{\partial_p}\overrightarrow{\partial}_{x}\right )}\\
\label{ZAB}\end{aligned}\end{equation}

An example of the fermionic system is given by that of the Dirac fermions with the action
\begin{equation}
S[\bar\psi,\psi]=\int d^4x \bar\psi(x)\hat D(\partial_x) \psi(x)
\label{Z} \end{equation}
where $\bar{\psi}$ and $\psi$ are the Dirac spinor fields. Depending on the nature of the given problem they may be understood either as complex - valued spinors or as operators or as the Grassmann - valued fields. Dirac operator is defined here as
\begin{equation}
\hat D(\partial_x) =i\gamma^{\mu}\partial_{\mu}-M
\end{equation}
Action may be written as
\begin{equation}
S[\bar\psi,\psi]=
\bra{\bar\psi}\hat D\ket{\psi}
\label{Z} \end{equation}
where we introduce shorthand notations $\bra{\bar\psi}$ and $\ket{\psi}$. Their meaning is  $\bra{\bar\psi}x\rangle = \bar{\psi}(x)$, and $\langle x \ket{\psi} = \psi(x)$.

\subsection{Relation between the Green function and Dirac operator in Weyl-Wigner formalism}

In continuous case we have
\begin{equation} \begin{aligned}
(i\gamma_\mu\partial^\mu_x-m)G(x-y)=\delta(x-y)
\label{green_func}\end{aligned} \end{equation}
which can be rewritten in the "operator" form
\begin{equation} \begin{aligned}
\bra{x} \hat D \hat G \ket{y}=\braket{x|y}
\label{DG}\end{aligned} \end{equation}
Applying Weyl-Wigner transformation, we obtain
\begin{equation} \begin{aligned}
(\hat Q \hat G)_W=Q_W\star G_W=1
\label{Gronewold}\end{aligned} \end{equation}
 Eq. (\ref{Gronewold}) is the Gronewold equation.


\subsection{Wilson fermions} \label{Wilson fermions}

One of the methods to discretize the Dirac field is Wilson fermions model (see Appendix \ref{AppB}) for more details. Action for Wilson fermions in Euclidean space discretized using rectangular lattice has the form
\begin{equation}\begin{aligned}
S^{(W)}_{F}=
\sum_{\substack{n,m\\ \alpha,\beta}} \hat{\bar\psi}_\alpha(n)D_{\alpha\beta}^{(W)}(n,m) \hat{\psi}_\beta(n)
\label{Wilson_Euclidean_action} \end{aligned} \end{equation}
where
\begin{equation}\begin{aligned}
D_{\alpha\beta}^{(W)}(n,m)=(\hat M+4)\delta_{nm}\delta_{\alpha\beta}-\frac{1}{2}\sum_\mu
\left[
(1-\gamma_\mu)_{\alpha\beta}\delta_{m,n+\hat\mu}+(1+\gamma_\mu)_{\alpha\beta}\delta_{m,n-\hat\mu}
\right]
\label{Z} \end{aligned} \end{equation}
Here lattice sites are referred to as $n,m$. $n+\hat{\mu}$ means the lattice site situated one lattice spacing ahead of $n$ in direction $\mu$. Indices $\alpha$ and $\beta$ correspond to Dirac matrices $\gamma_k$.
Inserting Fourier transform of the field
\begin{equation}
\psi ({r}_n)=\int_{\mathcal M} \frac{d^D {p}}{|\mathcal M|}e^{i{r}_n{p}}\psi ({p})
\label{4} \end{equation}
(where $|\mathcal M| = (2\pi)^D$ is volume of momentum space)
in the action (\ref{Wilson_Euclidean_action}) we obtain the terms like
\begin{equation} \begin{aligned}
&\sum_{{r}_n,{r}_m}
\bar \psi({r}_m) \delta_{{r}_n\pm{\bf e}_i,{r}_m}
\psi({r}_n)=
\sum_{{r}_n}
\bar \psi({r}_n\pm{\bf e}_i)
\psi({r}_n)=\\
& \sum_{{r}_n}
\int_{\mathcal M} \frac{d^D {p}}{|\mathcal M|}e^{-i({r}_n \pm {\bf e}_i){p}}\bar \psi ({p})
\int_{\mathcal M} \frac{d^D {q}}{|\mathcal M|}e^{i{r}_n{q}}\psi ({q})=
\int_{\mathcal M} \frac{d^D {p}}{|\mathcal M|}
\bar \psi ({p})
e^{\mp i{\bf e}_i{p}}
\psi ({p})
\label{I1_}
\end{aligned} \end{equation}
Here ${\bf e}_i$ is the unit lattice vector in the $i$ - th direction. Then, the action in momentum space becomes
\begin{equation}
S(\bar\psi,\psi)=\int \frac{d^D {p}}{|{\cal M}|}\bar\psi({p})  Q({p})\psi({p})
\label{action_lattice_p} \end{equation}
where
\begin{equation} \begin{aligned}
{Q}({p}) =\sum_{k=1,2,3,4} -i\gamma^k g_k ({p})+m({p})=
-i\Big[\sum_{k=1,2,3,4} \gamma^k g_k ({p})-im({p})\Big]=
\label{WF0}\end{aligned} \end{equation}
\begin{equation} \begin{aligned}
g_k({p})=\sin( p_k) \quad\quad m({p})=
m^{(0)}+\sum_{\nu=1}^4 (1-\cos(p_\nu))
\label{Z}\end{aligned} \end{equation}

\subsection{Wilson fermions in the presence of gauge field}
In continuous coordinates space, the transition from the Dirac operator in coordinates representation to momentum representation in presence of a gauge field $A$, is obvious.
\begin{equation}
S(\bar\psi,\psi)=\int \frac{d^D {p}}{|{\cal M}|}\bar\psi({p})  Q({p} - {A}(i\partial_{p}))\psi({p})
\label{Z} \end{equation}
In lattice theory it demands some work. In coordinates space, in the presence of gauge field, the Dirac operator takes the form
\begin{equation}
{D}_{\bf x,y}=-\frac{1}{2}\sum_i \left[ (1+\gamma^i)\delta_{{\bf x+e}_i,{\bf y}}+
(1-\gamma^i)\delta_{{\bf x-e}_i,{\bf y}}\right]U_{{\bf x},{\bf y}}+
(m^{(0)}+4)\delta_{{\bf x},{\bf y}}
\label{D} \end{equation}
while
\begin{equation}
U_{x,y}=Pe^{i\int_{ x}^{ y} d {\pmb \xi} { A}({\pmb \xi})}
\label{D} \end{equation}
We restrict ourselves by the case of the $U(1)$  gauge field $\bf A$ and then this parallel transporter is given by
\begin{equation}
U_{ x,y}=e^{i\int_{x}^{ y} d {\pmb \xi} { A}({\pmb \xi})}
\label{D} \end{equation}
Using the \textbf{Peierls substitution} the partition function
\begin{equation}
Z=\int D\bar{\psi}D\psi exp\left(-\sum_{{r}_n,{r}_m}\bar\psi({r}_m)\left({D}_{{r}_n,{r}_m}\right)\psi({r}_n)\right)
\label{Z} \end{equation}
may be written in the momentum representation

\begin{equation}
Z=\int D\bar{\psi}D\psi exp\left(\int \frac{d^D {p}}{|{\cal M}|}\bar\psi({p})  Q({p} - { A}(i\partial_{p}))\psi({p})\right)
\label{Z_Wilson} \end{equation}
In fact, the same refers to the other lattice models defined by operators $\hat{Q}$ different from that of the model of Wilson fermions.


\subsection{Approximate generalization of Weyl-Wigner fromalism from continuous space to lattice}
Definitions of phase space formalism for the continuous case (\ref{<A>}), (\ref{A_W}), (\ref{W_W}), (\ref{Moyal}) are modified for the case of discrete coordinates $x_n$ as follows,
\begin{equation} \begin{aligned}
&[\hat A]_W(x_n,p)=
\int_\mathcal{M}  dq e^{iqx_n} \bra{p+\frac{q}{2}} \hat{A} \ket{p-\frac{q}{2}}
\label{Z}\end{aligned} \end{equation}
\begin{equation} \begin{aligned}
&[\hat \rho]_W(x_n,p)=W(x,p)=
\int_\mathcal{M}  dq e^{-iqx_n} \bra{p-\frac{q}{2}} \hat{\rho} \ket{p+\frac{q}{2}}
\label{Z}\end{aligned} \end{equation}
\begin{equation}
\bra{\Psi} \hat{A} \ket{\Psi}=
\sum_{x_n} \int_\mathcal{M} \frac{dp}{\mathcal{M}} A_W(x_n,p) \rho_W(x_n,p)
\label{Z} \end{equation}
Where $\mathcal{M}$ is the first Brillouin zone and $x_n$ are the lattice points. We denote the trace of Weyl symbol as follows:
\begin{equation}
{\rm Tr} {A}_W =
\sum_{x_n} \int_\mathcal{M} \frac{dp}{\mathcal{M}} A_W(x_n,p)
\label{WeylTr} \end{equation}
\\
In one - dimensional case we will have
\begin{equation} \begin{aligned}
&[\hat A]_W(x_n,p)=
\int\limits_{-\frac{\pi}{a}}^{\frac{\pi}{a}}  dq e^{iqx_n} \bra{p+\frac{q}{2}} \hat{A} \ket{p-\frac{q}{2}}
\label{Z}\end{aligned} \end{equation}
Let us consider the Moyal product of the Weyl symbols of two operators in the one - dimensional case.
Weyl symbol of the product of two operators, called the Moyal product, is defined as follows (this time in momentum representation)
\begin{equation} \begin{aligned}
(\hat A \hat B)_W(x_n,p)&=
\int\limits_{-\pi/a}^{\pi/a}  dq e^{iqx_n} \bra{p+\frac{q}{2}} \hat{A} \hat B \ket{p-\frac{q}{2}}\\
&=
\int\limits_{-\pi/a}^{-\pi/a}  dq \int\limits_{-\pi/a}^{-\pi/a}  dk
e^{iqx_n} \bra{p+\frac{q}{2}}  \hat{A}
\ket{k}\bra{k} \hat B \ket{p-\frac{q}{2}}
\label{Z}\end{aligned} \end{equation}\\
changing variables
\begin{equation}\begin{aligned}
q=u+v \,\,\,\,\,\,\,\,\,\,\,\, k=p-u/2+v/2
\label{Z}\end{aligned}\end{equation}
\begin{equation}\begin{aligned}
u=\frac{1}{2}q-k+p \,\,\,\,\,\,\,\,\,\,\,\, v=\frac{1}{2}q+k-p
\label{Z}\end{aligned}\end{equation}
we come to
\begin{equation}\begin{aligned}
(\hat A \hat B)_W=&
\int\limits_{\substack{\text{Integration} \\ \text{area}}}  du \,\,dv \,\,
e^{iux} \bra{p+\frac{u}{2}+\frac{v}{2}}  \hat{A} \ket{p-\frac{u}{2}+\frac{v}{2}}
e^{ivx} \bra{p-\frac{u}{2}+\frac{v}{2}} \hat B \ket{p-\frac{u}{2}-\frac{v}{2}}
\label{Z}\end{aligned} \end{equation}
In case of the near diagonal operators, the only important region is around the origin, hence, we can change the integration area back to the square form\\
\begin{equation}\begin{aligned}
(\hat A \hat B)_W=&
\int\limits_{-\frac{\pi}{a}}^{\frac{\pi}{a}}  du
\int\limits_{-\frac{\pi}{a}}^{\frac{\pi}{a}} dv
e^{iux_n} \bra{p+\frac{u}{2}+\frac{v}{2}}  \hat{A} \ket{p-\frac{u}{2}+\frac{v}{2}}
e^{ivx_n} \bra{p-\frac{u}{2}+\frac{v}{2}} \hat B \ket{p-\frac{u}{2}-\frac{v}{2}}\\
=&\int\limits_{-\frac{\pi}{a}}^{\frac{\pi}{a}}  du
\int\limits_{-\frac{\pi}{a}}^{\frac{\pi}{a}}  dv
\left[e^{iux} \bra{p+\frac{u}{2}} \hat{A} \ket{p-\frac{u}{2}}\right]
e^{\frac{i}{2} \left( \overleftarrow{\partial_{x_n}}\overrightarrow{\partial_p}-\overleftarrow{\partial_p}\overrightarrow{\partial_{x_n}}\right )}
\left[e^{ivx} \bra{p+\frac{v}{2}} \hat B \ket{p-\frac{v}{2}}\right]
\label{Z}\end{aligned} \end{equation}
and get the same result, for the Moyal product, as for the continuous space
\begin{equation}\begin{aligned}
(AB)_W(x_n,p)\equiv A_W(x_n,p)\star B_W(x_n,p)=A_W(x_n,p)e^{\overleftrightarrow{\Delta}} B_W(x_n,p)
\label{Moyal}\end{aligned}\end{equation}
where
\begin{equation}\begin{aligned}
\overleftrightarrow{\Delta} \equiv{\frac{i}{2} \left( \overleftarrow{\partial}_{x_n}\overrightarrow{\partial_p}-\overleftarrow{\partial_p}\overrightarrow{\partial}_{x_n}\right )}\\
\label{ZAB}\end{aligned}\end{equation}
Although this expression has been derived for the case of one dimensional lattice, obviously it remains valid for the lattice models in any number of dimensions.

\subsection{Weyl - Wigner transform - general properties}\label{sec11}

Weyl symbol of operators introduced above establishes correspondence between operators and functions defined on phase space. It satisfies a certain set of properties typical for the constructions of deformational quantization. In fact, we may use the other definitions of Weyl symbol, which posses the same properties in order to explore various non - dissipative transport effects. Those basic properties are:
\begin{enumerate}
	\item{Star product identity}
	\be
	A_W(x,p) \star B_W(x,p) = (\hat A\hat B)_W(x,p).
	\label{*-def}
	\ee
	\item{First trace identity}
	\be
	\Tr {A_W} = \tr \hat A
	\label{Tr-def-1}
	\ee
	\item{Second trace identity}
	\be
	\Tr [A_W(x,p) \star B_W(x,p)] = \Tr [A_W(x,p) B_W(x,p)].
	\label{Tr-def-2}
	\ee
	\item{Weyl symbol of identity operator}
	\be
	({\hat 1})_W(x,p) = 1.
	\label{id-def}
	\ee
	\item{Star product}
	\be
	\star_{xp} \equiv e^{ \frac{i}{2} \(
		\ola{\pd_x}\,\ora{\pd_p}-
		\ola{\pd_p}\,\ora{\pd_x}
		\)}
	\label{star}
	\ee
\end{enumerate}
By $\rm tr$ we understand the trace of the operator itself in the original Hilbert space, while $\rm Tr$ is the trace operation defined for the Weyl symbols. In particular, for the approximate Wigner - Weyl calculus defined above it is given by Eq. (\ref{WeylTr}). \\ \\
Let us mention also the following useful property
\be
\left[ \hat A \hat B \hat C\right]_W=A_W \star B_W \star C_W
\ee
(See Appendix \ref{AppC}.)

It is worth mentioning, that the precise Wigner - Weyl calculus on the lattice has been proposed in \cite{FZ2020}. It obeys the above abstract properties precisely unlike the case of the introduced above approximate Wigner - Weyl calculus. This calculus has been designed for the models defined on rectangular lattices, and it allows to derive useful expressions for Hall conductivity to be discussed partially in the present paper as well. (Next we will use the similar constructions to investigate chiral separation effect.) It is important, however, that throughout the present paper we are limited to the case, when sums over lattice points in expressions for the total currents may be substituted by integrals. This is possible, when various inhomogeneities existing in the theory, are sufficiently small at the distances of the order of lattice spacings. Under these conditions the precise constructions of \cite{FZ2020} are not, in fact, necessary, and we are able to use the approximate Wigner - Weyl calculus for the lattice models described above.

\section{Electric current in Wigner - Weyl formalism}

\subsection{Partition function variation}

In this section we repeat briefly the considerations of \cite{ZW2019} that lead to the construction of topological expression for the quantum Hall conductivity of non - homogeneous systems. We will use later this technique to consider the non - homogeneous CSE.

In Euclidian space-time the partition function of a noninteracting fermionic system is expressed through the inverse bare Green function. We call it further for simplicity the Dirac operator and denote by  $\hat Q$. The partition function  is given by
\be
Z = \int D\bar{\psi }D\psi
\,\, e^{S[\psi ,\bar\psi  ]}
\label{Z01}
\ee
Here $\psi, \bar{\psi}$ are the Grassmann - valued fields, while $S$ is the action
\begin{equation}\begin{aligned}
&S[\psi ,\bar\psi  ]=\int_{\mathcal M} \frac{d^D{p}}{|\mathcal M|}\bar\psi({p}) \hat Q(i\partial_{p},{p})\psi({p})=\\
&\int_{\mathcal M} \frac{d^D{p}}{|\mathcal M|}\bar\psi^a({p}) \hat Q^{ab}(i\partial_{p},{p})\psi^b({p})=\\
&\sum_{{r}_n} \int_{\mathcal M} \frac{d^D{p}}{|\mathcal M|}  Q_W^{ab}({r}_n,{p}) W^{ba}({r}_n,{p})=\\
&\sum_{{r}_n} \int_{\mathcal M} \frac{d^D{p}}{|\mathcal M|}  \tr  \Big[ Q_W({r}_n,{p})  W({r}_n,{p})\Big]
\label{Z}\end{aligned}  \end{equation}
where we used Weyl symbols of operators
\be Q_W({r}_n,{p})=\hat Q_W \label{Q_W}\ee
\be W({r}_n,{p})=\(\ket{\psi}\bra{\psi}\)_W \ee
Here by $\ket{\psi}\bra{\psi}$ we denote operator with Grassmann - valued matrix elements
${\psi(x)}{\bar{\psi}}(y)$.
For simplicity of notations we discretize both space coordinates and imaginary time. This is usual for the lattice discretized relativistic field theory and unusual for the lattice models of condensed matter physics. In the latter case we are able to take off the discretization of imaginary time at any step of calculations in order to arrive at the conventional expression $\hat{Q} = i \omega - \hat{H}$, where $\hat{H}$ is one - particle Hamiltonian.

Using  \textit{Peierls} substitution \cite{KZ2017}, in the presence of gauge field  (\ref{Q_W}) takes the form
\be Q_W({p})\rightarrow Q_W({p}-{A}(i\pd_{p})) \ee
\begin{equation} \begin{aligned}
Z=\int D\bar\psi D\psi  \exp
\left(-\sum_{{r}_n} \int_{\mathcal M|} \frac{d^D{p}}{|\mathcal M|}  \tr  \Big[ Q_W({r}_n,{p})  W({r}_n,{p})\Big]\right)
\label{Z}\end{aligned} \end{equation}
Propagator of fermions is defined as
\begin{equation}
\hat G=-\frac{1}{Z}\int D\bar\psi D\psi \ket{\psi} \bra{\bar\psi} \exp\left(\int \frac{d^D{p}}{|\mathcal M|}\bar\psi({p}) \hat Q(i\partial_{p},{p})\psi({p})\right)
\label{Z} \end{equation}
Its expression in momentum space is
\begin{equation}\begin{aligned}
G({p}_1,{p}_2)=&\bra{{p}_1} G \ket{{p}_2}=\\
& \frac{1}{Z}\int D\bar\psi D\psi \bar\psi({p}_2) \psi({p}_1) \exp\left(\int \frac{d^D{p}}{|\mathcal M|}\bar\psi({p}) \hat Q(i\partial_{p},{p})\psi({p})\right)
\label{Z}\end{aligned}  \end{equation}
Variation of partition function may be expressed as follows
\begin{equation} \begin{aligned}
&\delta \log Z=\\
&-\frac{1}{Z}
\int D\bar\psi D\psi
\Big[\sum_{{r}_n} \int \frac{d^D{q}}{|\mathcal M|}
\delta Q^{ab}_W({r}_n,{q}) W^{ba}({r}_n,{q})\Big]
\exp \left(-\sum_{{r}_n} \int \frac{d^D{p}}{|\mathcal M|}  Q^{ab}_W({r}_n,{p})  W^{ba}({r}_n,{p})\right)=\\
&- \sum_{{r}_n} \int \frac{d^D{q}}{|\mathcal M|}
\delta Q^{ab}_W({r}_n,{q})
\left[ \frac{1}{Z} \int D\bar\psi D\psi  W^{ba}({r}_n,{q})
\exp \left(-\sum_{{r}_n} \int \frac{d^D{p}}{|\mathcal M|}  Q^{ab}_W({r}_n,{p})  W^{ba}({r}_n,{p})\right) \right]=\\
& \sum_{{r}_n} \int \frac{d^D{q}}{|\mathcal M|}
\delta Q^{ab}_W({r}_n,{q}) G^{ba}_W({r}_n,{q})=\\
& \sum_{{r}_n} \int \frac{d^D{q}}{|\mathcal M|}
\tr \left[ \delta Q_W({r}_n,{q}) G_W({r}_n,{q}) \right]
\label{Z}\end{aligned} \end{equation}
We obtain
\begin{equation} \begin{aligned}
&\delta \log Z= \sum_{{r}_n} \int \frac{d^D{p}}{|\mathcal M|}
\tr \left[ \delta Q_W({r}_n,{p}) G_W({r}_n,{p}) \right]
\label{dlogZ}\end{aligned} \end{equation}
that is
\begin{equation} \begin{aligned}
&\delta \log Z= \tr \left[\hat G \delta \hat Q  \right]=\Tr[G_W\star \delta Q_W]=
\Tr[ G_W\delta Q_W]
\label{Z}\end{aligned} \end{equation}
here we use definitions from section \ref{sec11}.\\
{
From now on we use continuum limit for the coordinates ${r}_n\rightarrow x$. This is possible if variations of fields on the distances of the order of lattice spacings are neglected. }\\
In the presence of gauge field we substitute $p\rightarrow p-A$
\be
Q_W(x,p)\rightarrow Q_W(x,p-A)
\ee
Variation with respect to the gauge field $A\rightarrow A+\delta A$ gives
\be
Q_W(x,p-(A+\delta A))=Q_W(x,p-A)+\pd_{A_i}Q_W(x,p-A)\delta A_i
\ee
and
\be
\delta Q_W=\pd_{A_i} Q_W \delta A_i=-\pd_{p_i} Q_W \delta A_i
\ee
Electric current is given by
\be
j_i(x)=\frac{\delta \log Z}{\delta A_k(x)}=
-\int_{\mathcal M} \frac{d^Dp}{|\mathcal M|}
\tr \left[  G_W(x,p) \partial_{p_i} Q_W(x,p)  \right]
\label{j_i}\ee

\subsection{Groenewold equation}
Dirac operator and Green function obey the following equation
\be \hat Q \hat G=1 \ee
Weyl-Wigner transform gives Groenewold equation
\be Q_W(p,x) \star  G_W(p,x)=1 \label{Groen}\ee
As a result of the variation with respect to the gauge field $A\rightarrow A+\delta A$

\be
\(Q_W+\delta Q_W\)\star \(G_W+\delta G_W\)\approx
Q_W\star G_W+\delta Q_W\star G_W+Q_W\star \delta G_W
\ee
hence
\be
\delta G_W=-G_W\star \delta Q_W \star G_W
\ee
\subsection{Topological invariance} \label{TI}
Integrating (or summing on the lattice) local current density of (\ref{j_i}), we obtain the total integrated current
\be
J_i&\equiv \int dx j_i(x)= -
\int dx \int_{\mathcal M} \frac{d^Dp}{|\mathcal M|}
\tr \left[  G_W(x,p) \partial_{p_i} Q_W(x,p)  \right]=\\
&=-\Tr \left[  G_W(x,p) \partial_{p_i} Q_W(x,p)  \right]
=-\Tr \left[  G_W(x,p) \star \partial_{p_i} Q_W(x,p)  \right]
\label{I_i}\ee
It is worth mentioning that $J$ is not the conventional current $I$, which is defined as an integral of current density over the cross section of a given sample. Relation between the two may be understood easily for the homogeneous system of rectangular form with length $L$ at finite temperature $1/\beta$. Then $J = \beta L I$. The last equality in Eq. (\ref{I_i}) is valid if spacial boundary conditions are periodic. For the variation of $J$ we obtain
\be
\delta J_i=\delta \Tr\left[G_W\star \partial_{p_i} Q_W\right]=
\Tr\left[
\delta G_W\star \partial_{p_i} Q_W+G_W\star \partial_{p_i} \delta Q_W
\right]
\ee
Using identity
\be
\delta G_W\star \partial_{p_i} Q_W
=-G_W\star \delta Q_W \star G_W\star \partial_{p_i} Q_W
\ee
and
\be
G_W\star \partial_{p_i} \delta Q_W
=\pd_{p_i}\(G_W\star \delta Q_W\)-
\pd_{p_i}G_W\star \delta Q_W
\ee
as well as
\be
\pd_{p_i} G_W=-G_W \star \(\pd_{p_i} Q_W\) \star G_W
\ee
and periodic boundary conditions in momentum space
we get
\be
\delta J_i=
- \Tr\left[
-G_W\star \delta Q_W \star G_W\star \partial_{p_i} Q_W+
G_W \star \(\pd_{p_i} Q_W\) \star G_W\star\delta Q_W
\right]
\ee
The cyclic properties of the trace will give
\be
\delta J_i=0
\ee
Hence, $J_i$ is topological invariant in the presence of periodic spacial boundary conditions.
Notice, that the above consideration fails in the presence of external electric field, when periodic boundary conditions cannot be imposed. Therefore, the appearance of non - vanishing response of $J$ to external electric field does not contradict with the statement that $J$ is topological invariant for the systems with periodic boundary conditions.

\subsection{Linear response}

Let us consider the case when the external gauge field $C$ is present
\be Q_W(p,x)=Q_W(p-C(x),x)\ee
We assume here that $Q_W$ has an additional space dependence to that coming from the gauge field. The gauge field itself is divided to the background one $B(x)$ and to that for which we are looking a linear response $A(x)$.\\
\be C(x)=A(x)+B(x)
\ee
%
Hence, the Dirac operator may be written as
\be Q_W(p,x)\approx Q_W^{(0)}(p,x)+\delta Q_W(p,x)
=Q_W^{(0)}(p,x)-\pd_{p_k}Q_W^{(0)}(p,x)A_k(x)\ee
where
\be Q_W^{(0)}(p,x)=Q_W^{(0)}\(p-B(x),x\) \ee
The propagator may also be presented as a perturbation
\be G_W(p,x)\approx G_W^{(0)}(p,x)+\delta G_W(p,x)\label{G_approx]}\ee
substituting this variation back into Groenewold equation (\ref{Groen}), and leaving only the first order, we get
\be
(Q_W^{(0)}+\delta Q_W)\star (G_W^{(0)}+\delta G_W) =
\underbrace{Q_W^{(0)}\star G_W^{(0)}}_{=1}+
Q_W^{(0)}\star \delta G_W+\delta Q_W\star G_W^{(0)}=1
\label{Groenwold_approx}\ee
\be \delta G_W=-G_W^{(0)}\star \delta Q_W\star G_W^{(0)}=
G_W^{(0)}\star \pd_{p_k}Q_W^{(0)}(p,x)A_k(x) \star G_W^{(0)}
\ee
using the identity (\ref{e^d}) we may write, up to the linear terms in $A_{ij}=\pd_iA_j-\pd_jA_i$:
\be
&\delta G_W(p,x)=\\
&\Big[G_W^{(0)}(p,x)\star
e^{-\frac{i}{2}\ola{\pd}_p \ora{\pd}_y}
\pd_{p_k}Q_W^{(0)}(p,x)A_k(y)
e^{\frac{i}{2}\ola{\pd}_y \ora{\pd}_p}
\star G_W^{(0)}(p,x)\Big]_{y=x}\approx \\
&\Big[G_W^{(0)}(p,x)\star
\(1-\frac{i}{2}\ola{\pd}_p \ora{\pd}_y\)
\pd_{p_k}Q_W^{(0)}(p,x)A_k(y)
\(1+\frac{i}{2}\ola{\pd}_y \ora{\pd}_p\)
\star G_W^{(0)}(p,x)\Big]_{y=x}\approx \\
&
\Big[G_W^{(0)}(p,x)\star \(\pd_{p_k}Q_W^{(0)}(p,x)\) \star G_W^{(0)}(p,x)\Big] A_k(x)-\\
&\frac{i}{2}\Big[\(\pd_{p_i}G_W^{(0)}(p,x)\)
\star \(\pd_{p_k}Q_W^{(0)}(p,x)\) \star G_W^{(0)}(p,x)\Big]
\pd_{x_i}A_k(x)+\\
&\frac{i}{2} \Big[ G_W^{(0)}(p,x) \star \(\pd_{p_k}Q_W^{(0)}(p,x)\) \star
\(\pd_{p_i}G_W^{(0)}(p,x)\) \Big] \pd_{x_i}A_k(x)=\\
&\Big[G_W^{(0)}\star \(\pd_{p_k}Q_W^{(0)}\) \star G_W^{(0)}\Big] A_k+
\frac{i}{2} \Big[ G_W^{(0)}\star \(\pd_{p_i} Q_W^{(0)}\) \star G_W^{(0)}
\star \(\pd_{p_j} Q_W^{(0)}\) \star G_W^{(0)} \Big]A_{ij}
\ee
we used above the following identity
\be
\pd_{p_i}\(G_W^{(0)} \star Q_W^{(0)}\)=0=
\(\pd_{p_i} G_W^{(0)} \)\star Q_W^{(0)}+
 G_W^{(0)} \star \(\pd_{p_i} Q_W^{(0)}\)
\label{d(GQ)}\ee
as well as
\be
\pd_{p_i} G_W^{(0)}=-G_W^{(0)} \star \(\pd_{p_i} Q_W^{(0)}\) \star G_W^{(0)}
\ee
Hence, we may represent $\delta G_W$ as follows
\be
\delta G_W(p,x)=&G_{W(k)}^{(1)}A_k+G_{W(ij)}^{(2)}A_{ij}=\\
&\Big[G_W^{(0)}\star \(\pd_{p_k}Q_W^{(0)}\) \star G_W^{(0)}\Big] A_k+\\
&\frac{i}{2} \Big[ G_W^{(0)}\star \(\pd_{p_i} Q_W^{(0)}\) \star G_W^{(0)}
\star \(\pd_{p_j} Q_W^{(0)}\) \star G_W^{(0)} \Big]A_{ij}
\label{deltaG}\ee
\be
G_W(p,x)\approx&G_W^{(0)}+G_{W(k)}^{(1)}A_k+G_{W(ij)}^{(2)}A_{ij}\\
\label{G=G0+G1+G3}\ee
\be
G_{W(i)}^{(1)}=
&\Big[G_W^{(0)}\star \(\pd_{p_i}Q_W^{(0)}\) \star G_W^{(0)}\Big] =
-\pd_{p_i}G_W^{(0)}\\
\label{G1}\ee
\be
G_{W(ij)}^{(2)}=
\frac{i}{2} \Big[ G_W^{(0)}\star \(\pd_{p_i} Q_W^{(0)}\) \star G_W^{(0)}
\star \(\pd_{p_j} Q_W^{(0)}\) \star G_W^{(0)} \Big]
\label{G2}\ee

\subsection{Electric current and gradient expansion}

In case of sufficiently weak inhomogeneity discussed above (\ref{dlogZ}) becomes
\begin{equation} \begin{aligned}
&\delta \log Z= \int dx \int_{\mathcal M} \frac{d^Dp}{|\mathcal M|}
\tr \left[ \delta Q_W(x,p) G_W(x,p) \right]
\label{Z}\end{aligned} \end{equation}
or, generally speaking
\begin{equation} \begin{aligned}
&\delta \log Z= \tr \left[\hat G \delta \hat Q  \right]=\Tr[G_W\star \delta Q_W]=
\Tr[ G_W\delta Q_W]
\label{Z}\end{aligned} \end{equation}
In the presence of gauge field
\begin{equation} \begin{aligned}
&\delta \log Z=-\int dx \int_{\mathcal M} \frac{d^Dp}{|\mathcal M|}
\tr \left[  G_W(x,p) \partial_{p_k} Q_W(x,p)  \right]\delta A_k(x)
\label{Z}\end{aligned} \end{equation}
As it was mentioned above the definition of current
\begin{equation} \begin{aligned}
&\delta \log Z \equiv \int dx j_k(x) \delta A_k(x)
\label{Z}\end{aligned} \end{equation}
gives
\be
j_k(x)=\frac{\delta \log Z}{\delta A_k(x)}=
-\int_{\mathcal M} \frac{d^Dp}{|\mathcal M|}
\tr \left[  G_W(x,p) \partial_{p_k} Q_W(x,p)  \right]
\label{jkx}\ee
{
The total integrated current (defined as an integral over space - time of the current density) is given by response to external uniform electric field
\be
J_k&=\int dx j_k(x)=-
\int dx
\int_{\mathcal M} \frac{d^Dp}{|\mathcal M|}
\tr \left[G_W(x,p) \partial_{p_k} Q_W(x,p)\right]\\
\ee

Using property (\ref{Tr-def-2}) (it is valid if periodic spacial boundary conditions are imposed), we obtain
\be
J_k&=-\Tr\left[G_W(x,p) \partial_{p_k} Q_W(x,p)\right]
=-\Tr\left[G_W(x,p)\star \partial_{p_k} Q_W(x,p)\right]
\ee
}
Here
\be
\pd_{p_i} Q_W=\pd_{p_i}Q_W^{(0)}-\(\pd_{p_i}\pd_{p_j}Q_W^{(0)}\)A_j
\ee
and
\be
&G_W\pd_{p_k} Q_W=\(G_W^{(0)}+G_{W(l)}^{(1)}A_l+G_{W(mn)}^{(2)}A_{mn}\)
\(\pd_{p_i}Q_W^{(0)}-\(\pd_{p_i}\pd_{p_j}Q_W^{(0)}\)A_j\)\\
\label{GdpQ}\ee
the first two terms in (\ref{GdpQ}) are
\be
&\(G_W^{(0)}+G_{W(l)}^{(1)}A_l\)\(\pd_{p_i}Q_W^{(0)}-\(\pd_{p_i}\pd_{p_j}Q_W^{(0)}\)A_j\)=\\
&G_W^{(0)}\pd_{p_i}Q_W^{(0)}-G_W^{(0)}\(\pd_{p_i}\pd_{p_j}Q_W^{(0)}\)A_j+
G_{W(l)}^{(1)}A_l\pd_{p_i}Q_W^{(0)}+O(A^2)
\ee
Since
\be
G_{W(j)}^{(1)}A_j\pd_{p_i}Q_W^{(0)}=-\(\pd_{p_j}G_W^{(0)}\)A_j\pd_{p_i}Q_W^{(0)}
\ee
up to the linear terms in $A$ we have
\be
&\(G_W^{(0)}+G_{W(l)}^{(1)}A_l\)\(\pd_{p_i}Q_W^{(0)}-
\(\pd_{p_i}\pd_{p_j}Q_W^{(0)}\)A_j\)=\\
&G_W^{(0)}\pd_{p_i}Q_W^{(0)}-
G_W^{(0)}\(\pd_{p_i}\pd_{p_j}Q_W^{(0)}\)A_j-
\(\pd_{p_j}G_W^{(0)}\)\(\pd_{p_i}Q_W^{(0)}\)A_j=\\
&G_W^{(0)}\pd_{p_i}Q_W^{(0)}-
\pd_{p_j}\(G_W^{(0)}\pd_{p_i}Q_W^{(0)}\)A_j
\ee
The third term in (\ref{GdpQ}) gives (we keep only the linear terms in $A_{ij}$):
\be
&G_{W(kl)}^{(2)}A_{kl}\(\pd_{p_i}Q_W^{(0)}-\(\pd_{p_i}\pd_{p_j}Q_W^{(0)}\)A_j\)\approx
G_{W(kl)}^{(2)} \pd_{p_i}Q_W^{(0)}A_{kl}=\\
&\frac{i}{2} \Big[ G_W^{(0)}\star \(\pd_{p_k} Q_W^{(0)}\) \star G_W^{(0)}
\star \(\pd_{p_l} Q_W^{(0)}\) \star G_W^{(0)} \Big] \pd_{p_i}Q_W^{(0)} A_{kl}
\ee
Writing current of Eq. (\ref{jkx}) in terms of $A_k$ and $A_{mn}$, we get
\be
j_i(x)= -
\int_{\mathcal M} \frac{d^Dp}{|\mathcal M|}
\tr \Bigg[&
G_W^{(0)}\pd_{p_i}Q_W^{(0)}-\\
&\pd_{p_k}\(G_W^{(0)}\pd_{p_i}Q_W^{(0)}\)A_k+\\
&\frac{i}{2} \Big[ G_W^{(0)}\star \(\pd_{p_m} Q_W^{(0)}\) \star G_W^{(0)}
\star \(\pd_{p_n} Q_W^{(0)}\) \star G_W^{(0)} \Big] \pd_{p_i}Q_W^{(0)} A_{mn}
\Bigg]
\label{jkx1}\ee
We may define
\be
j_i(x)=j_i^{(0)}(x)+j_{i(k)}^{(1)}(x)A_k(x)+j_{i(mn)}^{(2)}(x)A_{mn}(x)
\ee
where
\be
j_i^{(0)}(x)=-
\int_{\mathcal M} \frac{d^Dp}{|\mathcal M|}
\tr \Bigg[&
G_W^{(0)}\pd_{p_i}Q_W^{(0)}
\Bigg]
\label{j0}\ee
and
\be
j_{i(k)}^{(1)}(x)=
\int_{\mathcal M} \frac{d^Dp}{|\mathcal M|}
\tr \Bigg[
&\pd_{p_k}\(G_W^{(0)}\pd_{p_i}Q_W^{(0)}\)
\Bigg]
\label{j1}\ee
in case of periodic boundary conditions in momentum space $j_{i(k)}^{(1)}(x)=0$ since it is a total derivative while
\be
j_{i(mn)}^{(2)}(x)=-
\int_{\mathcal M} \frac{d^Dp}{|\mathcal M|}
\tr \Bigg[
\frac{i}{2} \Big[ G_W^{(0)}\star \(\pd_{p_m} Q_W^{(0)}\) \star G_W^{(0)}
\star \(\pd_{p_n} Q_W^{(0)}\) \star G_W^{(0)} \Big] \pd_{p_i}Q_W^{(0)}
\Bigg]
\label{j2}\ee
The $j_{i(mn)}^{(2)}(x)$ is the local electric conductivity tensor since it is a coefficient in front of electromagnetic tensor.\\
The average electric conductivity (we  assume $A_{ij}=const$) is to be obtained from
\be
\bar{J}_i^{(2)}= J_i^{(2)}/V^{(4)}&\equiv\frac{1}{V^{(4)}}A_{mn} \int d^Dx j_{i(mn)}^{(2)}(x)=
\frac{1}{\beta {\bf V}}A_{mn} \int dx j_{i(mn)}^{(2)}(x)\\
&=-\frac{iA_{mn}}{2\beta {\bf V}}\int d^Dx \int_{\mathcal M} \frac{d^Dp}{|\mathcal M|}
\tr \Bigg[
\Bigg( G_W^{(0)}\star \(\pd_{p_m} Q_W^{(0)}\) \star G_W^{(0)}
\star \(\pd_{p_n} Q_W^{(0)}\) \star G_W^{(0)} \Bigg) \pd_{p_i}Q_W^{(0)}
\Bigg]\\
&=-\frac{iA_{mn}}{2\beta {\bf V}}\Tr
\Bigg[
\Bigg( G_W^{(0)}\star \(\pd_{p_m} Q_W^{(0)}\) \star G_W^{(0)}
\star \(\pd_{p_n} Q_W^{(0)}\) \star G_W^{(0)} \Bigg) \pd_{p_i}Q_W^{(0)}
\Bigg]
=\mathcal{W}_{mni}A_{mn}
\ee
where $V^{(4)}$ is the overall volume of Euclidean space - time while $\bf V$ is the three - dimensional volume.
We defined
\be
\mathcal{W}_{mni}\equiv
-\frac{i}{2\beta {\bf V}}
\Tr \Bigg[
\Bigg( G_W^{(0)}\star \(\pd_{p_m} Q_W^{(0)}\) \star G_W^{(0)}
\star \(\pd_{p_n} Q_W^{(0)}\) \star G_W^{(0)} \Bigg) \pd_{p_i}Q_W^{(0)}
\Bigg]
\ee
Thus the term in electric current giving response to external field strength is
\be
\bar{J}_i=\mathcal{W}_{mni} F_{mn}
\label{JWF}\ee

\subsection{Hall conductance }

Using expressions of Appendix A we obtain in the presence of external electric field
\be J_i=\sigma_{ij}E_j \ee
Using (\ref{JWF}) and (\ref{F^E}) we get
\be J_i=\mathcal{W}_{4ji} F^E_{4j}=
\frac{\mathcal{W}_{4ji}}{i}E_j\ee
We can define an averaged Hall conductivity
\be
\sigma_{mn}\equiv \frac{\mathcal{W}_{4[ji]}}{i}=
\frac{1}{i}\(\mathcal{W}_{4mn}-\mathcal{W}_{4nm}\)
\ee
For the $3+1$ D systems it may be rewritten as follows \cite{ZW2019}:
\begin{eqnarray}
\sigma_{kj} =  \frac{1}{2\pi^2}\epsilon^{kjl4} {\cal N}_{l} \label{calM0C},\, {\cal N}_l =  - \frac{T\epsilon_{ijkl}}{{\bf V}\, 3!\,8\pi^2}\ \int d^4 x d^4p \,{\rm Tr}\,  \Big[  {G}_W(p,x) \star \frac{\partial {Q}_W(p,x)} {\partial p_i}\star \frac{\partial  {G}_W(p,x)}{\partial p_j} \star \frac{\partial  {Q}_W(p,x)}{\partial p_k}  \Big]\label{rez1} \end{eqnarray}

\section{Chiral separation effect}

\subsection{Axial current}

Now we are equipped by all necessary tools to study chiral separation effect in non - homogeneous systems. This is an appearance of axial current in the fermionic systems with finite chemical potential and external magnetic field. The latter is supposed to be uniform, but the non - homogeneity of an arbitrary nature is present in the system even when the external magnetic field is off. It is assumed that the lattice model is similar somehow to the model of Wilson fermions - its "Dirac operator" $\hat Q$ is a $4\times 4 $ matrix expressed through the gamma matrices. However, the form of $\hat Q$ not necessarily repeats that of the Wilson Dirac operator. Moreover, the considered operators $\hat Q$ are in general not diagonal with respect to momentum.
The local axial current density may be defined as
\be
j^5_k(x)= -
\int_{\mathcal M} \frac{d^Dp}{|\mathcal M|}
\tr \left[ \gamma^5 G_W(x,p) \pd_{p_k} Q_W(x,p)  \right]
\label{ji5x}\ee
Repeating all steps of the previous section we come to the following term containing the linear response to external field strength:
\be
j_k^5(x)=-
\frac{i}{2}
\int_{\mathcal M} \frac{d^Dp}{|\mathcal M|}
\tr \Bigg[\gamma^5
 \Big[ G_W^{(0)}\star \(\pd_{p_i} Q_W^{(0)}\) \star G_W^{(0)}
\star \(\pd_{p_j} Q_W^{(0)}\) \star G_W^{(0)} \Big] \pd_{p_k}Q_W^{(0)}
\Bigg]
F_{ij}
\label{ji5lr}\ee
Integrating (or summing on the lattice) the local current given in (\ref{ji5x}), we get
\be
J_i^5&\equiv \int d^Dx j_i^5(x)=
-\int d^Dx \int_{\mathcal M} \frac{d^Dp}{{\bf v}|\mathcal M|}
\tr \left[\gamma^5  G_W(x,p) \partial_{p_i} Q_W(x,p)  \right]=\\
&=-\Tr \left[ \gamma^5 G_W(x,p) \partial_{p_i} Q_W(x,p)  \right]
=-\Tr \left[ \gamma^5 G_W(x,p) \star \partial_{p_i} Q_W(x,p)  \right]
\label{Ii5}\ee
Here $\bf v$ is volume of the lattice cell. We have a useful formula ${\bf v}|\mathcal M| = (2\pi)^D$.
Dividing by the total $4$ - volume we obtain the average axial current
\be
\bar{J}_k^5=\frac{J_k^5}{\beta{\bf V}}=&
-\frac{i}{2}\frac{1}{\beta{\bf V}}
\int d^Dx\int_{\mathcal M} \frac{d^Dp}{(2\pi)^D}\\
&\tr \Bigg[\gamma^5
\Big[ G_W^{(0)}\star \(\pd_{p_i} Q_W^{(0)}\) \star G_W^{(0)}
\star \(\pd_{p_j} Q_W^{(0)}\) \star G_W^{(0)} \Big] \pd_{p_k}Q_W^{(0)}
\Bigg]
F_{ij}
\label{Ji5lr}\ee


\subsection{(The absence of) topological invariance for the total axial current}

\label{top5}

Integrating over all space - time (or summing over the lattice) the local current density of (\ref{ji5x}), we get the total integrated axial current
\be
J_i^5&\equiv \int d^Dx j_i^5(x)=-
\int d^Dx \int_{\mathcal M} \frac{d^Dp}{(2\pi)^D}
\tr \left[\gamma^5  G_W(x,p) \partial_{p_i} Q_W(x,p)  \right]=\\
&=-\Tr \left[ \gamma^5 G_W(x,p) \partial_{p_i} Q_W(x,p)  \right]
=-\Tr \left[ \gamma^5 G_W(x,p) \star \partial_{p_i} Q_W(x,p)  \right]
\label{Ii5_}\ee
As for the case of electric current, the total integrated axial current differs from the conventional total current $I^5$ through the cross - section of the sample. For example, for the case of a uniform rectangular sample in the case when nothing depends on time we have $J^5 = \beta L I^5$, where $\beta = 1/T$ is inverse temperature (assumed to be large) while $L$ is the length of the sample. Variation of $J^5$ is given by
\be
\delta J_i^5=-\delta \Tr\left[\gamma^5 G_W\star \partial_{p_i} Q_W\right]=-
\Tr\left[
\gamma^5\delta G_W\star \partial_{p_i} Q_W+
\gamma^5G_W\star \partial_{p_i} \delta Q_W
\right]
\ee
Using identities
\be
\gamma^5\delta G_W\star \partial_{p_i} Q_W
=-\gamma^5 G_W\star \delta Q_W \star G_W\star \partial_{p_i} Q_W
\ee
and
\be
\gamma^5 G_W\star \partial_{p_i} \delta Q_W
=\gamma^5 \pd_{p_i}\(G_W\star \delta Q_W\)-
\gamma^5 \pd_{p_i}G_W\star \delta Q_W
\ee
as well as
\be
\pd_{p_i} G_W=-G_W \star \(\pd_{p_i} Q_W\) \star G_W
\ee
we obtain (for the case of periodic boundary conditions) that under the trace we may substitute
\be
\gamma^5 G_W\star \partial_{p_i} \delta Q_W
= \gamma^5 G_W \star \(\pd_{p_i} Q_W\) \star G_W\star\delta Q_W
\ee
As a result we come to
\be
\delta J_i= -
\Tr\left[
-\gamma^5 G_W\star \delta Q_W \star G_W\star \partial_{p_i} Q_W+
\gamma^5 G_W \star \(\pd_{p_i} Q_W\) \star G_W\star\delta Q_W
\right]
\ee
If $\gamma^5$ commutes (or anti - commutes) with $G$ and $Q$ then
\be
&\Tr\left[
\gamma^5 G_W\star \delta Q_W \star G_W\star \partial_{p_i} Q_W
\right]=\\
&\Tr\left[
\partial_{p_i} Q_W \star \gamma^5 G_W\star \delta Q_W \star G_W
\right]=\\
&\Tr\left[
G_W \star \partial_{p_i} Q_W \star \gamma^5 G_W\star \delta Q_W
\right]=\\
&\Tr\left[
\gamma^5 G_W \star \partial_{p_i} Q_W \star  G_W\star \delta Q_W
\right]
\ee
The latter condition means that the model possesses precise chiral symmetry.  Under this (very restrictive) condition we obtain
\be
\delta J_i^5=0
\ee
In the above considerations we also implied that there are no singularities of the Green function. The latter condition means that the fermions are gapped, and the Fermi energy is within the gap.
We conclude that for the systems with gapped fermions in the presence of precise chiral symmetry the total integrated axial current $J_i^5$ would be a topological invariant. In practise, however, the corresponding requirements are too restrictive. For example, for lattice Dirac fermions the presence of a gap (mass) excludes chiral symmetry. Therefore, in practise the total axial current cannot be topological invariant unlike the total electric current. (Recall that for the latter we need periodic boundary conditions, which exclude, in particular the presence of external electric field.) Below we will see, that the topological invariance ever appears in the consideration of axial currents of realistic systems in the form of robustness of the response of axial current to external magnetic field and chemical potential.

\subsection{Axial current for gapless fermions at finite temperature}

We are going to regularize the theory by finite (but small) temperature in order to deal with gapless fermions.
Matsubara frequencies are
$
p_4=\omega_n=\frac{2\pi\(n+\frac{1}{2}\)}{\beta}
$.
Here inverse temperature $\beta = 1/T$ is taken in lattice units:
$
N_t\equiv\frac{1}{T}
$,
and the values of $p_4$ are
$
p_4=\frac{2\pi\(n_4+\frac{1}{2}\)}{N_t}$, $ n_4=-\frac{N_t}{2},..,\frac{N_t}{2}-1
$.
The boundary values are
$
\om_{n=-\frac{N_t}{2}}=\frac{2\pi\(-\frac{N_t}{2}+\frac{1}{2}\)}{N_t}=
-\pi+\frac{\pi}{N_t}
$ and
$
\om_{n=\frac{N_t}{2}-1}=\frac{2\pi\(\frac{N_t}{2}-\frac{1}{2}\)}{N_t}=
\pi-\frac{\pi}{N_t}
$.
The Matsubara frequencies most close to zero are:
$
\om_{n=0}=\frac{\pi}{N_t}
$ and
$
\om_{n=-1}=-\frac{\pi}{N_t}
$.
One can see that $\om_n$ never equals to zero. Therefore, even for the system of massless/gapless fermions the propagator never has poles in momentum space. As a result the expression for the axial current is well - defined
\be
\bar{J}_k^5=&-
\frac{i}{2}\frac{1}{\beta{\bf V}}
\sum_{n=-\frac{N_t}{2}}^{\frac{N_t}{2}-1}
\int d^3x \int_{\mathcal M_3} \frac{d^3p}{(2\pi)^3}\\
&\tr \Bigg[\gamma^5
\Big[ G_W^{(0)}\star \(\pd_{p_i} Q_W^{(0)}\) \star G_W^{(0)}
\star \(\pd_{p_j} Q_W^{(0)}\) \star G_W^{(0)} \Big] \pd_{p_k}Q_W^{(0)}
\Bigg]
F_{ij}
\label{Ji5}\ee

Introducing the chemical potential $\omega_n\ra\omega_n-i\mu$
we obtain
\be
\bar{J}_k^5=&-
\frac{1}{{2\bf V}\beta}
\sum_{n=-\frac{N_t}{2}}^{\frac{N_t}{2}-1}
\int d^3x \int_{\mathcal M_3} \frac{d^3p}{(2\pi)^3}\\
&\pd_{\omega_n}\tr \Bigg[\gamma^5
\Big[ G_W^{(0)}\star \(\pd_{p_i} Q_W^{(0)}\) \star G_W^{(0)}
\star \(\pd_{p_j} Q_W^{(0)}\) \star G_W^{(0)} \Big] \pd_{p_k}Q_W^{(0)}
\Bigg]
F_{ij}\mu
\label{Ji5mu1}\ee
Here $|\mathcal M_3|$ is volume of the three - dimensional Brillouin zone. We represent the above expression as
\be
\bar{J}_k^5(x)=\mathcal{\sigma}_{ijk}F_{ij}\mu
\label{Ji5mu2}\ee
where
\be
\mathcal{\sigma}_{ijk}=&
-\frac{1}{{2\bf V}\beta}
\sum_{n=-\frac{N_t}{2}}^{\frac{N_t}{2}-1}
\int d^3x \int_{\mathcal M_3} \frac{d^3p}{(2\pi)^3}\\
&\pd_{\omega_n}\tr \Bigg[\gamma^5
\Big[ G_W^{(0)}\star \(\pd_{p_{[i}} Q_W^{(0)}\) \star G_W^{(0)}
\star \(\pd_{p_{j]}} Q_W^{(0)}\) \star G_W^{(0)} \Big] \pd_{p_k}Q_W^{(0)}
\Bigg]
\label{Nijk5}\ee
has the meaning of the CSE conductivity when external field strength corresponds to a constant magnetic field $H$: $F_{ij} = -\epsilon_{ijk} H_k$. Then
$$
\bar{J}_k^5(x)=-\mathcal{\sigma}_{ijk}\epsilon_{ijk^\prime} H_{k^\prime}\mu
$$
Here we assume antisymmetrization with respect to indices $i$ and $j$.
We will see below that for the wide range of systems $-\epsilon_{ijk} \sigma_{ijk^\prime} = \delta^{k k^\prime} \sigma_{CSE}$ with a scalar CSE conductivity.
We represent expression for the CSE conductivity as
\be
\mathcal{\sigma}_{ijk}=&
\sum_{n=-\frac{N_t}{2}}^{\frac{N_t}{2}-1}
\pd_{\om_n}\mathcal{\sigma}_{ijk}^{(3)}
\label{Nijk5_1}\ee
where
\be
\mathcal{\sigma}_{ijk}^{(3)}=&-
\frac{1}{{2\bf V}}
\int d^3x \int_{\mathcal M_3} \frac{d^3p}{(2\pi)^3}\\
&\tr \Bigg[\gamma^5
\Big[ G_W^{(0)}\star \(\pd_{p_{[i}} Q_W^{(0)}\) \star G_W^{(0)}
\star \(\pd_{p_{j]}} Q_W^{(0)}\) \star G_W^{(0)} \Big] \pd_{p_k}Q_W^{(0)}
\Bigg]
\label{N_3}\ee

\subsection{The limit of small temperature and CSE conductivity }

The limit of small temperature $T\ra 0$, $N_t\ra \infty$, $\frac{\pi}{N_t}=\ep\ra 0$ allows to replace the sum by an integral. However, the point $\omega = 0$ is excluded from this integral due to the above mentioned properties of finite temperature theory:
\be
\sum_{n=-\frac{N_t}{2}}^{\frac{N_t}{2}-1} \tab\ra \tab
\frac{\beta}{2\pi}\int_{-\pi+\ep}^{0-\ep}d\omega+ \frac{\beta}{2\pi}\int_{0+\ep}^{\pi-\ep}d\omega
\ee
Then (\ref{Nijk5}) becomes
\be
\mathcal{\sigma}_{ijk}&=
\lim_{\ep\ra0}
\int_{-\pi+\ep}^{0-\ep}d\om
\pd_{\om}\mathcal{\sigma}_{ijk}^{(3)}
+
\int_{0+\ep}^{\pi-\ep} d\om
\pd_{\om}\mathcal{\sigma}_{ijk}^{(3)}\\
&=\lim_{\ep\ra0}\[
\mathcal{\sigma}_{ijk}^{(3)}(-\pi+\ep) - \mathcal{\sigma}_{ijk}^{(3)}(0-\ep)+
\mathcal{\sigma}_{ijk}^{(3)}(0+\ep) - \mathcal{\sigma}_{ijk}^{(3)}(\pi-\ep)\]
\label{Nijk_int}\ee
using periodicity
\be
\mathcal{\sigma}_{ijk}^{(3)}(-\pi)=\mathcal{\sigma}_{ijk}^{(3)}(\pi)
\ee
we obtain
\be
&\mathcal{\sigma}_{ijk}=
\lim_{\ep\ra0}\[
\mathcal{\sigma}_{ijk}^{(3)}(0+\ep) +\(-\mathcal{\sigma}_{ijk}^{(3)}(0-\ep)\) \]
\label{Nijk_N_3}\\
\ee
where
\be
&{\sigma}_{ijk}^{(3)}(\om=0\pm \ep)=\\
&-\frac{1}{{2\bf V}}
\int d^3x \int_{\mathcal M_3} \frac{d^3p}{(2\pi)^4}
\tr \Bigg[\gamma^5
\Big[ G_W^{(0)}\star \(\pd_{p_{[i}} Q_W^{(0)}\) \star G_W^{(0)}
\star \(\pd_{p_{j]}} Q_W^{(0)}\) \star G_W^{(0)} \Big] \pd_{p_k}Q_W^{(0)}
\Bigg]\Bigg|_{\om=0\pm \ep}=\\
&-\frac{1}{{2\bf V}}
\int_{\mathcal M_3} \frac{d^3p}{(2\pi)^4}
\int d^3x
\tr \Bigg[\gamma^5
\Big[ G_W^{(0)}\star \(\pd_{p_{[i}} Q_W^{(0)}\) \star G_W^{(0)}
\star \(\pd_{p_{j]}} Q_W^{(0)}\) \star G_W^{(0)} \Big] \pd_{p_k}Q_W^{(0)}
\Bigg]\Bigg|_{\om=0\pm \ep}
\label{N_3_1}\ee

In the static case both the Green function $G$ and its inverse $Q$ do not depend on time. Therefore all possible singularities of the above expressions are situated at $\omega = 0$. Our integrals avoid these singularities due to the small but finite values of $\epsilon$. In the absence of inhomogeneity (when the stars may be omitted in the above expressions) at $\omega = 0$ the singularities of expressions standing in the integrals mark positions of Fermi surfaces.
The presence of inhomogeneity changes positions of those singularities. However, weak inhomogeneity cannot force those singularities to approach boundary of the Brillouin zone. On the language of effective low energy continuum theory of our lattice model we say that the inhomogeneity cannot force singularities of the Green functions and their products to approach infinity.

\subsection{CSE conductivity as a topological invariant}

In Eq. (\ref{Nijk_N_3}) the integrals entering $\sigma$ cancel each other except those in the small vicinities of the mentioned above singularities. That's why we may restrict integrations in Eq. (\ref{N_3_1}) by the small regions of the Brillouin zone above/below the singularities.  Here, in this region we assume the presence of precise chiral symmetry, which means that the effective low energy theory of our lattice model is chiral invariant (if the chiral anomaly is ignored). Recall, that the chiral symmetry cannot be maintained in the whole Brillouin zone of the majority of physical models. This is why the chiral anomaly appears. In the expression of Eq. (\ref{N_3_1}), however, we restrict integrations to the region, where $\gamma^5$ commutes/anti - commutes with $Q$ and $G$. We will see below that as a result the sum of the integrals in Eq. (\ref{Nijk_N_3}) represents a topological invariant, which does not depend on the form of the surface in $4D$ momentum space surrounding the singularities. We may deform this surface arbitrarily in such a way that it remains surrounding the singularities. This way instead of the two pieces of the infinitely close planes (situated above and below the singularities) we may integrate over the sphere in momentum space (this is illustrated by the Figure).

\begin{figure}[h]
	\centering  %
	\includegraphics{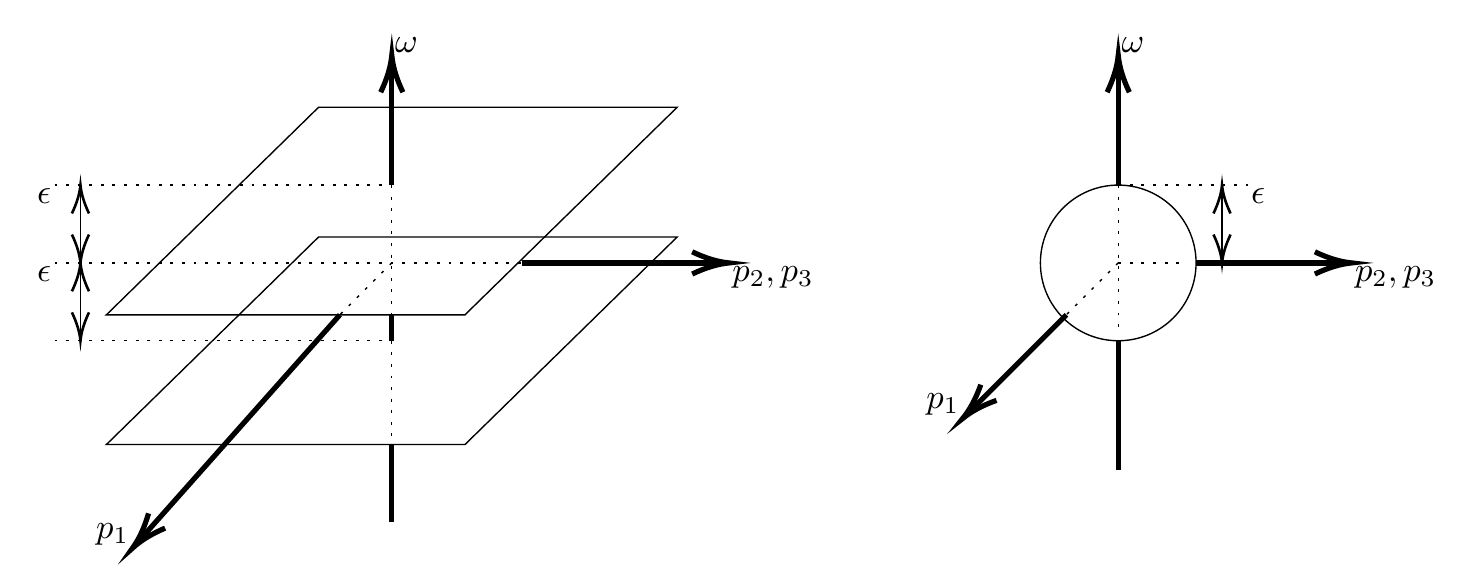}  %
	\caption{Deformation of the surface surrounding singularities of Eq. (\ref{Ncompl}). }  %
	\label{FIG}   %
\end{figure}

Thus we rewrite
\be
&\mathcal{\sigma}_{ijk}=\\
&-\frac{1}{{2\bf V}}
\int_{\Sigma_3} \frac{d^3p}{(2\pi)^4}
\int d^3x
\tr \Bigg[\gamma^5
\Big[ G_W^{(0)}\star \(\pd_{p_{[i}} Q_W^{(0)}\) \star G_W^{(0)}
\star \(\pd_{p_{j]}} Q_W^{(0)}\) \star G_W^{(0)} \Big] \pd_{p_k}Q_W^{(0)}
\Bigg]
\ee
Here the integral is over $\Sigma_3$, which is the 3D hypersurface in 4D momentum space. It consists of the two infinitely close pieces of the planes situated above and below the singularities of expression standing in the integral. Since $\gamma^5$ commutes/anti - commutes with $G$ and $Q$ in this region, we may rewrite this expression as
$$
\sigma_{ijk} = \epsilon_{ijk} \sigma_H/2
$$
where
\begin{equation}
\sigma_H = \frac{\mathcal{N}}{2\pi^2}\label{sigmaH}
\end{equation}
and
\begin{eqnarray}
\mathcal{N}&=&-\frac{\epsilon_{ijk}}{48 \pi^2 {\bf V}}
\int_{\Sigma_3} {d^3p}
\int d^3x
\tr \Bigg[\gamma^5
\Big[ G_W^{(0)}\star \(\pd_{p_{i}} Q_W^{(0)}\) \star G_W^{(0)}
\star \(\pd_{p_{j}} Q_W^{(0)}\) \star G_W^{(0)} \Big] \pd_{p_k}Q_W^{(0)}
\Bigg]=\nonumber\\
&=&-\frac{1}{48 \pi^2 {\bf V}}
\int_{\Sigma_3}
\int d^3x
\tr \Bigg[\gamma^5
G_W^{(0)}\star d Q_W^{(0)} \star G_W^{(0)}
\wedge \star d Q_W^{(0)}\star G_W^{(0)} \star \wedge d Q_W^{(0)}
\Bigg]\label{Ncompl}
\end{eqnarray}
This expression is topological invariant provided that $\gamma^5$ commutes or anti - commutes with $Q_W$ and $G_W$ in the vicinity of $\Sigma_3$. This may be proved easily using methods of Sect. \ref{top5}. Therefore, we may deform the surface $\Sigma_3$ in such a way that the deformed $\Sigma_3$ does not cross the singularities of expression standing inside the integral.  As it has been mentioned above, we may deform the surface, for example, in such a way that it will have the form of a sphere (this is illustrated by Fig. \ref{FIG}).

\subsection{The limit of a homogeneous system and calculation of CSE conductivity}

When space inhomogeneities are sufficiently weak we are able to omit the star product in the above expression for the CSE conductivity (see Appendix \ref{AppE}). This simplifies considerably calculation of $\sigma_{CSE}$. Our expression is then reduced to that of \cite{KZ2017}:
\begin{eqnarray}
&&\mathcal{N}=-\frac{1}{48 \pi^2 {\bf V}}
\int_{\Sigma_3}
\int d^3x
\tr \Bigg[\gamma^5
 G_W^{(0)} d Q_W^{(0)}   G_W^{(0)}
\wedge  d Q_W^{(0)}  G_W^{(0)} \wedge d Q_W^{(0)}
\Bigg]\label{Nsimple}
\end{eqnarray}
Let us discuss for the definiteness example of the system with Wilson fermions in the presence of weakly varying external electric potential (see Appendix \ref{AppB}). We assume that this model is used for the description of the continuous field theory with one massless fermion. This means that parameter $m^{(0)}$ is set to zero. In the absence of electric potential the model has one Fermi point at $p=0$. Calculation of Eq. (\ref{Nsimple}) in this case has been given in Appendix \ref{AppF}. It gives $\mathcal{N} = 1$. In the system of $N$ fermions of this type we obtain $\mathcal{N} = N$. The presence of electric potential, which is much smaller than the inverse lattice spacing, does not break chiral invariance of effective low energy theory. At the same time, it turns Fermi point into a Fermi surface. Nevertheless, as far as the introduced electric potential is weak enough, the value of $\mathcal{N}$ remains equal to that of the trivial homogeneous model. It is still given by Eq. (\ref{Nsimple}), in which $\Sigma_3$ embraces the Fermi surface.

Suppose that we modify functions $g_i$  and $m$ entering Eq. (\ref{Wilson_Q}) in such a way that they become depending on a new parameter of the dimension of length $l_g$. Suppose that $p l_g$ is of the order of unity around the singularities of expression under the integral of Eq. (\ref{Ncompl}).  In this case we already cannot omit star products in Eq. (\ref{Ncompl}). However the CSE conductivity still remains unchanged if modification of functions $g_i$ and $m$ is continuous. It is given by Eq. (\ref{Ncompl}), in which $\Sigma_3$ embraces all singularities of an expression of the corresponding integral. These singularities already do not repeat the positions of Fermi surfaces but are supposed not to approach infinity (or the inverse lattice spacing).

We come to an interesting conclusion, that the CSE conductivity remains unchanged if we modify the fermionic system under consideration. There are only two requirements to this modification: 1) it is smooth; 2) it does not break chiral invariance of an effective low energy theory. This observation gives us the simple receipt how to calculate in practise the CSE conductivity for the given system. We should find the simple homogeneous system like that of the Wilson fermions (or that of overlap fermions \cite{KZ2017}), which is connected to the given one by a continuous deformation. The Fermi surface is allowed to change its form during such a deformation. We should also require that the chiral symmetry is not broken in effective low energy theory neither in the original inhomogeneous system nor during the deformation to the mentioned simple homogeneous system. Finally, value of the CSE conductivity in the original complicated system is equal to its value in the simple one.

\section{Conclusions and discussions}
\label{SectConcl}

In the present paper we discuss Chiral Separation Effect in essentially nonhomogeneous systems of chiral fermions. This is an appearance of non - dissipative axial current along the direction of external magnetic field. Our main result is expression for the CSE conductivity, i.e. for the response of axial current to external magnetic field and to chemical potential (both are assumed to be independent of time and space coordinates):
\begin{equation}
\sigma_H = \frac{\mathcal{N}}{2\pi^2}\label{sigmaHC}
\end{equation}
with
\begin{eqnarray}
\mathcal{N}
&=&-\frac{1}{48 \pi^2 {\bf V}}
\int_{\Sigma_3}
\int d^3x
\tr \Bigg[\gamma^5 G_W^{}\star d Q_W^{}  \star G_W^{}
\wedge \star d Q_W^{} \star G_W^{} \star \wedge d Q_W^{}
\Bigg]\label{NcomplC}
\end{eqnarray}
Here $G_W$ is the Wigner transformed two point Green function, while $Q_W$ is Weyl symbol of lattice Dirac operator. Both are taken in the absence of external magnetic field. Both are matrices $4\times 4$, which may model systems of Dirac fermions in the lattice regularized QFT or in the condensed matter systems with emergent Dirac/Weyl fermions. We assume that the inhomogeneities of the system under consideration are negligible at the distance of the order of lattice spacing. In this expression the integral is taken along the closed surface $\Sigma_3$ that surrounds positions $\Xi(x)$ of all singularities of expression standing inside the integral. In the case of weak inhomogeneities (when the Moyal (star) products may be replaced by the ordinary ones) $\Xi(x)$ at each $x$ coincides with the position of the coordinate dependent Fermi surface. In a more general case the position of $\Xi(x)$ cannot be predicted easily, and it does not,  in general case, coincide with the position $\Xi^\prime(x)$ of the singularities of $G_W$. One may consider surface in phase space $\Xi$ (or surface $\Xi^\prime$) as an extension of the notion of Fermi surface to the case of nonhomogeneous system.  In turn, Eq. (\ref{NcomplC}) is an extension of the topological invariant $N_3$ responsible for the topological stability of the Fermi points/Fermi surfaces in relativistic quantum field theory \cite{Volovik2003} to the case of non-homogeneous systems. In both cases chiral symmetry remains essential for the topological stability of these objects.

We accept assumption that along $\Xi(x)$ at each $x$ matrix $\gamma^5$ commutes or anti - commutes with $Q_W$. This reflects the requirement that the considered fermions are chiral in the low energy limit - close to the zeros of $Q_W$ or poles of $G_W$. Under this condition expression of Eq. (\ref{NcomplC}) is a topoloigical invariant. It is robust to the smooth deformation of the system as long as $\Sigma_3$ does not cross $\Xi(x)$ for any $x$. This property of $\mathcal{N}$ allows us to make a strong statement about the CSE conductivity of nonhomogeneous system. It remains non - sensitive to the particular form of lattice Dirac operator $\hat{Q}$. To be explicit, we may start consideration from the system with lattice Wilson fermions in the presence of weak external electric potential. In such a system the value of $\mathcal{N}$ can be easily calculated and is equal to the number of the species of Wilson fermions (see Appendix \ref{AppF}). In fact, the same value of $\mathcal{N}$ (equal to the number of effective continuum Dirac fermions) can also be obtained for the other lattice regularizations (say, for overlap fermions) in the presence of slowly varying external fields. The details of calculations then repeat those of \cite{KZ2017}.  Next, we consider deformation of the system. It may result, for example, from elastic deformations in case of solid state systems or, say, from external gravitational field or from rotation in case of lattice regularized relativistic QFT. There may be the other reasons, which lead to modification of lattice Dirac operator $\hat{Q}$. We assume that under such deformations the chiral fermions remain chiral. This means that matrix $\gamma^5$ remains commuting/anti - commuting with ${Q}_W$ at low energies (i.e. in the vicinity of $\Xi(x)$). Surface $\Sigma_3$ should remain embracing $\Xi(x)$ for all values of $x$. Besides, it remains in the region of the Brillouin zone, where $\gamma^5$ remains commuting/anti - commuting with ${Q}_W$. The positions of $\Sigma_3$, $\Xi$ and $\Xi^\prime$ are assumed to remain in the small region of Brillouin zone. The size of this region is to be much smaller than the size of the Brillouin zone itself. In the effective low energy theory this region becomes the vicinity of zero momentum. On the language of effective low energy theory we require that the positions of singularities of the Green function do not approach infinity. This requirement is especially natural for the lattice regularized non - homogenous relativistic quantum field theory of chiral fermions. Recall that an ordinary Fermi surface of a homogeneous system cannot approach infinity. It is natural to suppose that $\Xi$ being an extension of the notion of Fermi surface to the case of non - homogeneous systems, also cannot approach infinity.

Thus we come to an interesting conclusion. {\it For any system of chiral fermions the axial current in the presence of constant external magnetic field is proportional to magnetic field. The coefficient of proportionality is equal to ${\rm const} + \sigma_H \, \mu$, where $\mu$ is chemical potential. Coefficient $\sigma_H$ is universal. Irrespective of the particular form of the system it is equal to $\frac{N}{2\pi^2}$, where $N$ is the number of species of chiral Dirac fermions. The positions and forms  of Fermi surfaces (and their extensions to the non - homogeneous systems) is irrelevant as long as the chiral symmetry is maintained in their small vicinities.} Recall that we cannot provide the precise chiral symmetry in the whole Brillouin zone because of the Nielsen - Ninomiya theorem except for the marginal case when $\sigma_H=0$. It is worth mentioning that the number of chiral Dirac fermions is not always equal to the number of components of the fermion field divided by $4$. For example, let us consider the model with Wilson fermions of Appendix \ref{AppB}. In this model at $m^{(0)}=0$ there is precisely one chiral Dirac fermion, $\mathcal{N}=4/4 = 1$. At the same time for $m^{(0)} = -2$ we will have $4$ chiral Dirac fermions, and ${\mathcal N}=4$. This is not possible to deform the model with $m^{(0)} = -2$ to the model with $m^{(0)}=0$ smoothly keeping chiral symmetry in small vicinity of the Fermi surface. We conclude, that the systems with $m^{(0)}=0$ and $m^{(0)}=-2$ belong to different homotopic classes of the models describing chiral fermions. Of course, $\mathcal{N}$, being the topological invariant in the class of chiral theories,  has different values for $m^{(0)}=0,-2$.

It is worth mentioning that in the above consideration we ignored completely effect of interactions between the fermions. We expect, however, that the topological expression of Eq. (\ref{NcomplC}) remains valid in the presence of interactions if we replace the noninteracting two - point Green function $\hat{G}$  by the complete interacting Green function. (The same refers to its inverse $\hat{Q}$.) This expectation is based on the recent consideration of the similar question for the quantum Hall effect of systems with interactions (see \cite{Zhang_Zubkov_JETP_2019}).  Notice, that radiative corrections to the CSE in QED calculated in \cite{Shovkovy} contain singularities in the limit of vanishing electron or photon mass. This reflects the well - known problem of  infrared/collinear singularities in the systems with massless fermions interacting via an exchange by gauge bosons. For this reason in high energy physics the interacting fermions are typically considered with finite mass. In the case of Weyl and Dirac semimetals, however, the emergent Dirac fermions are true massless. Instead of the exchange by photons we are to consider Coulomb interactions. We do expect that the topological nature of CSE conductivity for chiral fermions survives in the presence of these interactions. Infrared divergencies are to be treated in this case carefully. However, the explicit consideration of this question remains out of the scope of the present paper.

We expect wide applications of results obtained here both in condensed matter theory and in relativistic high energy physics. In the latter case the chiral separation effect of chiral fermions is specific for the quark gluon plasma. The quark gluon plasma appears, in particular, during the heavy ion collisions. The fireballs existing just after a collision are in the presence of strong external magnetic field. Axial current of the CSE results after the decay of the fireball in an asymmetry of the outgoing particles. This asymmetry has been observed in experiment giving an evidence of the CSE. At the same time, all previous theoretical descriptions of this phenomenon dealt with the simplified homogeneous systems. This description is, of course, far from reality. In practise the system of chiral quarks inside the fireball is highly inhomogeneous. From the very beginning it was not clear which kind of CSE effect we may observe in this case. Our present answer is very simple - the complicated structure of the system does not affect {\it at all} the value of CSE conductivity $\sigma_H$. It remains equal to its value from the homogeneous model. In order to prove this statement, we need to assume, though, that the external magnetic field is homogeneous. This requirement is not realistic as well. But overall, the pattern of the CSE that follows from the above study is much more close to reality than the one that follows from the consideration of naive homogeneous systems.

In condensed matter physics the analogue of the chiral separation effect emerges, first of all, in an effective description of fermionic superfluid $^3He-A$. In this system the chiral fermions appear in vicinities of the Fermi points. The superfluid component of liquid may exist in the non - homogeneous state (say, forming various vortices) thus giving rise to the non - homogeneous effective  theory of fermionic quasiparticles. We expect that indirectly the results obtained above may be extended to this effective theory. The direct extension is not possible here because atoms of $^3He$ are neutral, and therefore, only the emergent $U(1)$ gauge field appears in this case. This emergent gauge field is axial rather than vector. And this fact complicates an analogy.

The direct observation of chiral separation effect may be performed for the Dirac and Weyl semimetals, where emergent relativistic chiral Dirac fermions appear in the vicinity of the Fermi points. Here we may apply constant external magnetic field and consider the axial current of the CSE that appears due to the finite chemical potential. Again, the realistic systems are not homogeneous because of the presence of impurities, because of dislocations and disclinations of the crystals, and because of elastic deformations. This is an ideal setup for the observation of the CSE of inhomogeneous systems, where, as we expect, our findings may become an important ingredient of its description. It is worth mentioning that there is a certain technical difficulty related to the experimental detection of the axial current in Dirac/Weyl semimetals. It is not as simple as the detection of electric current or even spin current. However, we hope that the future development of experimental technique will resolve this difficulty.

\section*{Acknowledgements}
The authors kindly acknowledge numerous discussions with I.Fialkovsky, C.X.Zhang and Xi Wu.

\appendix

\section{Tensor conventions in Minkowski and Euclidean spaces}

\label{AppA}

In 4D Minkowski space with metric $\eta_{\mu\nu}=(+---)$, where ${\mu,\nu}=0,1,2,3$, we have
{
\be
&x^\mu_M=(t,\vec x) \tab x^0_M=t \tab x^i_M=(\vec x)_i
\\
&x_\mu^M=(t,-\vec x) \tab x_0^M=t \tab x_i^M=-(\vec x)_i
\ee
and
\be
&p^\mu_M=(E,\vec p)\tab p^0_M=E \tab p^i_M=(\vec p)_i
\\
&p_\mu^M=(E,-\vec p) \tab p_0^M=E \tab p_i^M=-(\vec p)_i
\ee
Therefore,
\be
\pd_i^M=\frac{\pd}{\pd x^i_M}=\(\pd_t,\vec \nabla\)\\
\pd^i_M=\frac{\pd}{\pd x_i^M}=\(\pd_t,-\vec \nabla\)
\ee
Conventional quantum mechanical operators are:
\be
&\hat p^0_M=\hat p_0^M=i\pd_t=i\pd_0^M=i\pd^0_M\\
&\hat p^i_M=-i\nabla_i=i\pd^i_M\\
&\hat p_i^M=i\nabla_i=i\pd_i^M
\ee
and
\be
\hat p^\mu_M=i\pd^\mu_M \tab \hat p_\mu^M=i\pd_\mu^M
\ee
In 4D Euclidean space metric is $\eta_{\mu\nu}=(++++)$, where $\mu,\nu=1,2,3,4$. Therefore,
\be
x^\mu_E=x_\mu^E=(\vec x, ix_0) \tab  x_i^E=x^i_M=(\vec x)_i \tab x_4^E=ix_0^M=it
\ee
and
\be
\pd_\mu^E=\frac{\pd}{\pd x_\mu^E}
\ee
Correspondingly,
\be
&\pd_i^E=\frac{\pd}{\pd x^i_M}=\pd_i^M=\vec \nabla_i
\\
&\pd_4^E=\frac{\pd}{\pd x^4}=\frac{\pd}{\pd ix^0}=-i\pd_0^M=-i\pd_t
\ee
We may choose definitions of conventional quantum mechanical operators as follows
 $\hat p^E_\mu=i\pd^E_\mu$
\be
&\hat p^E_i=i\pd^E_i=\hat p_i^M=-\hat p^i_M=-\(\hat {\vec p}\)_i\\
&\hat p^E_4=i\pd^E_4=i\(-i\pd_0\)=-i\hat p_0
\ee
and
\be
x_\mu^E p_\mu^E=x_i^E p_i^E+x_4^E p_4^E=-\vec x \cdot \vec p+x_0p_0=x_\mu^M p^\mu_M
\ee
Alternatively, we may define
$\hat p^E_\mu=-i\pd^E_\mu$
\be
&\hat p^E_i=-i\pd^E_i=-\hat p_i^M=\hat p^i_M=\(\hat {\vec p}\)_i\\
&\hat p^E_4=-i\pd^E_4=-i\(-i\pd_0\)=i\hat p_0
\ee
and
\be
x_\mu^E p_\mu^E=x_i^E p_i^E+x_4^E p_4^E=\vec x \cdot \vec p-x_0p_0=-x_\mu^M p^\mu_M
\ee
}
Therefore,
\be
F^{i0}=\pd^iA^0-\pd^0A^i=
-\pd_iA^0-\pd^0A^i=
-(\nabla\phi)_i-\pd_t(\vec{A})_i=E^i
\ee
and
\be
F_{0i}=\pd_0A_i-\pd_iA_0=\pd_t(-\vec{A})_i-(\nabla\phi)_i=E_i
\ee
In Euclidean space with $x_4=ix_0$, $\pd_4=-i\pd_0$ we
 choose  $\hat p^E_\mu=i\pd^E_\mu$. Hence
\be
p_i^E=-\(\vec p\)_i \tab p_4=-ip_0\\
A_i^E=-\(\vec A\)_i\tab A_4=-iA_0
\ee
and
\be
E_i=F_{0i}^M
&=\pd_0^M A_i^M-\pd_i^M A_0^M\\
&=\pd_0^M\(A_i^E\)-\pd_i^M\(iA_4^E\)\\
&=i\pd_4^E\(A_i^E\)-\pd_i^E\(iA_4^E\)\\
&=i(\pd_4A_i^E-\pd_iA_4^E)=iF_{4i}^E
\ee
Thus we come to the following relation between Euclidean field strength and real electric field of Minkowski space
\be F^E_{4i}=-iE_i \label{F^E}\ee

\section{Wilson fermions}

\label{AppB}
Action for the Wilson fermions (defined on rectangular lattice) is
\begin{equation}\begin{aligned}
S^{(W)}_{F}=
\sum_{\substack{n,m\\ \alpha,\beta}} \hat{\bar\psi}_\alpha(n)K_{\alpha\beta}^{(W)}(n,m) \hat{\psi}_\beta(n)
\label{Wilson_Euclidean_action} \end{aligned} \end{equation}
where
\begin{equation}\begin{aligned}
K_{\alpha\beta}^{(W)}(n,m)=(\hat M+4)\delta_{nm}\delta_{\alpha\beta}-\frac{1}{2}\sum_\mu
\left[
(1-\gamma_\mu)_{\alpha\beta}\delta_{m,n+\hat\mu}+(1+\gamma_\mu)_{\alpha\beta}\delta_{m,n-\hat\mu}
\right]
\label{Z} \end{aligned} \end{equation}
Here indices $m, n$ enumerate lattice points while $\alpha, \beta$ are spinor indices.
By $n+\hat\mu$ we denote shift of the lattice point by one link in the $\mu$ - th direction.
Let us calculate the Fourier transform
\be
K_{\alpha\beta}(nm)=
\int_{-\pi}^\pi \frac{d^4p}{(2\pi)^4}
\tilde K_{\alpha\beta}(p) e^{ip(n-m)}
\label{K_Fur}\ee
Lattice Dirac operator of Wilson fermions is then defined as
\be
Q_{\alpha\beta}(p)\equiv \tilde K_{\alpha\beta}(p)=
\left[
\sum_{k=1,2,3,4} i\gamma_k g_k (p)+m(p)
\right]_{\alpha\beta}
=
i\left[\sum_{k=1,2,3,4} \gamma_k g_k (p)-im(p)\right]_{\alpha\beta}\label{Wilson_Q}
\ee
where
\begin{equation} \begin{aligned}
g_k(p)=\sin( p_k) \quad\quad m(p)=
m^{(0)}+\sum_{\nu=1}^4 (1-\cos(p_\nu))
\label{Z}\end{aligned} \end{equation}
The two-point function
\be
G_{\alpha\beta}(n,m)\equiv K^{-1}_{\alpha\beta}(n,m)=
-\braket{\hat\psi_\alpha(n)\hat{\bar\psi}_\beta(m)}
\ee
The inverse matrix is defined by
\be
\sum_{\lambda,l}
K^{-1}_{\alpha\lambda}(n,l)
K_{\lambda\beta}(l,m)=
\delta_{\alpha\beta}\delta_{nm}
\label{sum_K-1_K}\ee
Then in four - dimensional space
\be
K^{-1}_{\alpha\beta}(n,m)=
\int_{-\pi}^\pi \frac{d^4p}{(2\pi)^4}
G_{\alpha\beta}(p)e^{ip(n-m)}
\label{K-1}\ee
Inserting (\ref{K-1}) and (\ref{K_Fur}) into (\ref{sum_K-1_K}), we get
\be
\sum_{\lambda,l}
\int_{-\pi}^\pi \frac{d^4p}{(2\pi)^4}
G_{\alpha\lambda}(p)e^{ip(n-l)}
\int_{-\pi}^\pi \frac{d^4k}{(2\pi)^4}
\hat K_{\lambda\beta}(k)e^{ik(l-m)}
=
\delta_{\alpha\beta}\delta_{nm}
\label{sum_2}\ee
and
\be
\int_{-\pi}^\pi \frac{d^4p}{(2\pi)^4}
\sum_{\lambda}
G_{\alpha\lambda}(p)
\hat K_{\lambda\beta}(p)
e^{ip(n-m)}
=
\delta_{\alpha\beta}
\int_{-\pi}^\pi \frac{d^4p}{(2\pi)^4} e^{ip(n-m)}
\label{sum_3}\ee
Then
\be
\sum_{\lambda}
G_{\alpha\lambda}(p)
\hat K_{\lambda\beta}(p)=\delta_{\alpha\beta}
\ee
gives
\be
&\(\sum_{k=1,2,3,4} i\gamma_k g_k (p)+m(p)\)_{\alpha\lambda}
\(\sum_{q=1,2,3,4} -i\gamma_q g_q (p)+m(p)\)_{\lambda\beta}\\
&=\delta_{\alpha\beta}
\(\sum_{k=1,2,3,4}g_k^2(p)+m^2(p)\)
\label{QQ}\ee
Hence
\be
G_{\alpha\beta}(p)=
\frac{\left[\sum_{q} -i\gamma_q g_q (p)+m(p)\right]_{\alpha\beta}}
{\sum_{k}g_k^2(p)+m^2(p)}
\label{G_p}
\ee
In the presence of electromagnetic field, $p\ra p-A(i\pd_p)$, using Peierls formula, the Dirac operator for Wilson fermions may be represented as $Q(p)\ra Q(p-A(i\pd_p)) $
\be
Q_{\alpha\beta}(p-A(i\pd_p))=
\left[
\sum_{k} i\gamma_k g_k (p-A(i\pd_p))+m(p-A(i\pd_p))
\right]_{\alpha\beta}
\ee
or, in the operator manner:
\be
Q_{\alpha\beta}(\hat p-A(\hat x))=
\left[
\sum_{k} i\gamma_k g_k (\hat p-A(\hat x))+m(\hat p-A(\hat x))
\right]_{\alpha\beta}
\ee
We are interested to find the inverse of $Q$ in this case in the same way as in (\ref{QQ}).
It can be shown that the contribution due to the commutation relations $[\hat p, A(\hat x)]$ is the one lattice spacing shift to each harmonic. So, as long as the space dependence of the field is much larger than lattice spacing, we may neglect this contribution.
Hence, the Green's function in the presence of gauge field under the conditions described above (we use (\ref{G_p})) takes the form
{
\be
G_{\alpha\beta}(p-A(i\pd_p))\approx
\frac{\left[\sum_{q} -i\gamma_q g_q (p-A(i\pd_p))+m(p-A(i\pd_p))\right]_{\alpha\beta}}
{\sum_{k}g_k^2(p-A(i\pd_p))+m^2(p-A(i\pd_p))}
\ee
}
in the operator notations (and matrix form) we obtain
\be
\hat G(p-A(i\pd_p))\approx
\frac{\left[\sum_{q} -i\gamma_q g_q (p-A(i\pd_p))+m(p-A(i\pd_p))\right]_{\alpha\beta}}
{\sum_{k}g_k^2(p-A(i\pd_p))+m^2(p-A(i\pd_p))}
\ee

\section{Translation operator and the Moyal product properties}

\label{AppC}
Let us consider translation operator
\be e^{a\pd_x}f(x)=f(x+a)\ee
\be e^{a\pd_x}\(f(x)g(x)\)=f(x+a)g(x+a)=\(e^{a\pd_x}f(x)\)\((e^{a\pd_x}g(x))\)\ee
One can prove this identity as follows.
The left hand side has the form
\be e^{a\pd_x}\(f(x)g(x)\)&=
\sum_{n=0}^\infty\frac{1}{n!}\pd_x^n \(f(x)g(x)\)=
\sum_{n=0}^\infty\frac{1}{n!} \sum_{k=0}^n \frac{n!}{(n-k)!k!} f^{(n-k)}g^{(k)}
\ee
The right hand side is
\be &\(e^{a\pd_x}f(x) \) \(e^{a\pd_x}g(x)\)=
\(\sum_{n=0}^\infty\frac{1}{n!}\pd_x^n f(x)\)
\(\sum_{k=0}^\infty\frac{1}{k!}\pd_x^k g(x)\)\\
&=\frac{1}{0!}f^{(0)} \frac{1}{0!}g^{(0)}\\
&+\frac{1}{1!}f^{(1)}\frac{1}{0!}g^{(0)}+
\frac{1}{0!}f^{(0)}\frac{1}{1!}g^{(1)}\\
&+\frac{1}{2!}f^{(2)}\frac{1}{0!}g^{(0)}+
\frac{1}{1!}f^{(1)}\frac{1}{1!}g^{(1)}+
\frac{1}{0!}f^{(0)}\frac{1}{2!}g^{(2)}\\
&+\frac{1}{3!}f^{(3)}\frac{1}{0!}g^{(0)}+
\frac{1}{2!}f^{(2)}\frac{1}{1!}g^{(1)}+
\frac{1}{1!}f^{(1)}\frac{1}{2!}g^{(2)}+
\frac{1}{0!}f^{(0)}\frac{1}{3!}g^{(3)}...\\
&=\sum_{n=0}^\infty\sum_{k=0}^n \frac{1}{(n-k)!k!} f^{(n-k)}g^{(k)}
\ee
Hence
\be e^{a\pd_x}\(f(x)g(x)\)=\(e^{a\pd_x}f(x)\)\((e^{a\pd_x}g(x))\)\label{e^d}\ee
We may represent Moyal product of Weyl symbols of three operators as follows
\be \(\hat A \hat B \hat C\)_W(x,p)=
\Big(A_W(x,p)\star B_W(x,p)\Big)\star C_W(x,p)=
A_W(x,p)\star \Big(B_W(x,p)\star C_W(x,p)\Big)
\ee
that is
\be
&\Big(f_1(x,p)\star f_2(x,p)\Big)\star f_3(x,p)
= \(f_1(x,p)
e^{ \frac{i}{2} \(
\ola{\pd_x}\,\ora{\pd_p}-
\ola{\pd_p}\,\ora{\pd_x}
\)}
f_2(x,p)\)
e^{ \frac{i}{2} \(
	\ola{\pd_x}\,\ora{\pd_p}-
	\ola{\pd_p}\,\ora{\pd_x}
	\)}
f_3(x,p)\\
&=\left[e^{\frac{i}{2}\(
\pd_{x_1}\pd_{p_2}-\pd_{p_1}\pd_{x_2}+
\pd_{x_1}\pd_{p_3}+\pd_{x_2}\pd_{p_3}-
\pd_{p_1}\pd_{x_3}-\pd_{p_2}\pd_{x_3}
\)}
f_1(x_1,p_1)f_2(x_2,p_2)f_3(x_3,p_3)
\right]_{\substack{x_1=x_2=x_3=x\\p_1=p_2=p_3=p}}\\
&=
\left[
e^{\frac{i}{2}\(
	\pd_{x_1}(\pd_{p_2}+\pd_{p_3})+
	\pd_{x_2}(\pd_{p_3}-\pd_{p_1})+
	\pd_{x_3}(-\pd_{p_2}-\pd_{p_1})
	\)}
f_1(x_1,p_1)f_2(x_2,p_2)f_3(x_3,p_3)
\right]
_{\substack{x_1=x_2=x_3=x\\p_1=p_2=p_3=p}}
\ee

\section{Model based on Wilson fermions. Conditions for the elimination of star products.}

\label{AppE}

Expression
\be
\Bigg[\gamma^5
\Big[ G_W^{(0)}\star \(\pd_{p_i} Q_W^{(0)}\) \star G_W^{(0)}
\star \(\pd_{p_j} Q_W^{(0)}\) \star G_W^{(0)} \Big] \pd_{p_k}Q_W^{(0)}
\Bigg]\label{trexample}
\ee
includes the terms like
\be
g_i \star g_j
\ee
where
\be \star=
e^{\frac{i}{2}\(\ola{\pd_x}\,\ora{\pd_p}-\ola{\pd_p}\,\ora{\pd_x} \)}=
1+\frac{i}{2}\(\ola{\pd_x}\,\ora{\pd_p}-\ola{\pd_p}\,\ora{\pd_x} \)+...
\ee
Hence
\be
g_i \star g_j=  g_i  g_j +
g_i
\frac{i}{2}\(\ola{\pd_x}\,\ora{\pd_p}-\ola{\pd_p}\,\ora{\pd_x} \)
g_j +...
\ee
\be
\Big[\pd_x g_i\Big(pa-A(x)a\Big) \Big]=
-\pd_{p_i} g_i \pd_xA_i(x)=ag_i'\pd_xA_i(x)
\ee
\be
\Big[\pd_p g_j\Big(pa-A(x)a\Big) \Big]=
ag_j'
\ee
Therefore,
\be
\Big[\pd_xg_i\Big]\Big[\pd_pg_j\Big]=
g_i' g_j' a^2 \pd_xA_i(x)
\ee
Since
\be
-1\le g_i, g_i' \le 1
\ee
if
\be
a^2 \pd_x A_i(x)<<1
\ee
then the star products may be replaced by the ordinary ones.

The physical meaning of the above statement may be easily understood if we will consider the sinusoidal space dependence $A(x)\sim {\rm sin}(2\pi \, x/\lambda)$. Then
\be
a^2\frac{A}{\la}<<1 \tab \tab aA<<\frac{\la}{a}
\ee
Under these conditions in the above expression of Eq. (\ref{trexample}) we may omit the star.

\section{Calculation of topological invariant responsible for the  CSE conductivity for the Wilson fermions in the presence of slowly varying external fields}
\label{AppF}

It has been shown in Appendix \ref{AppE} that for sufficiently weak inhomogeneity for the model of Wilson fermions the Moyal products may be replaced by the ordinary ones in the expression for Hall conductivity. The condition for this is $a^2\frac{A}{\la}<<1$, where $a$ is the lattice spacing, $A$ is the external electromagnetic potential while $\lambda$ is the typical wavelength of external field $A$. Therefore, the singularities of expression standing in the integral in this case are placed along the singularities of the (Wigner transformed) Green functions. We have
\be
Q(\hat p-A(\hat x))=
\left[
\sum_{k} i\gamma_k g_k (\hat p-A(\hat x))+m(\hat p-A(\hat x))
\right]
\ee
and
\be
\hat G(\^p-A(\^x))=
\frac{\left[\sum_{q} -i\gamma_q g_q (\^p-A(\^x))+m(\^p-A(\^x))\right]}
{\sum_{k}g_k^2(\^p-A(\^x))+m^2(\^p-A(\^x))}
\ee
where
\be
g_k(p)=\sin( \^p_k)
\ee
\be
m(\^p)
&=m^{(0)}+\sum_{k=1}^4 (1-\cos(\^p_k))\\
&=m^{(0)}+\sum_{k=1}^4 2\sin^2\(\frac{\^p}{2}\)\\
&=m^{(0)}+\sum_{k=1}^4 2g_k^2\(\frac{\^p}{2}\)
\ee
The Green's function poles are given by the solutions of equation
\be
\sum_{k}g_k^2(\^p-A(\^x))+m^2(\^p-A(\^x))=0
\ee
For the massless fermions, when $m^{(0)}=0$,
\be
Q(\hat p-A(\hat x))=
\sum_{k} \Bigg[ i\gamma_k g_k (\hat p-A(\hat x))+2g_k^2 \(\frac{\hat p-A(\hat x)}{2}\)
\Bigg]
\ee
and
\be
\hat G(\^p-A(\^x))=
\frac{\left[\sum_{q} -i\gamma_q g_q (\^p-A(\^x))+2g_q^2 \(\frac{\hat p-A(\hat x)}{2}\)\right]}
{\sum_{k}g_k^2(\^p-A(\^x))+4\(\sum_j g_j^2\(\frac{\hat p-A(\hat x)}{2}\)\)^2} \equiv \frac{R(p-A)}{U(p-A)}
\ee
we have
\be
R=&\sum_{k=1}^3\[-i\ga_k\sin(p_k-A_k)+2\sin^2\(\frac{p_k-A_k}{2}\)\]-\\
&-i\ga_4\sin(p_4-A_4)+\sin^2\(\frac{p_4-A_4}{2}\)
\ee
and
\be
U=&\sum_{k=1}^3\sin^2(p_k-A_k)+\sin^2(p_4-A_4)+\\
&+4\[\sum_{k=1}^3\sin^2\(\frac{p_k-A_k}{2}\)+\sin^2\(\frac{p_4-A_4}{2}\)\]^2
\ee
The poles are space dependent and for $A_4 = i\phi = 0$ they correspond to the single points
\be
p_i=A_i(x)
\ee
In the neighborhood of these points
\be
p_i-A_i(x)=\xi_i\ra 0 \tab\ra\tab g_i(\xi)\approx\xi_i
\ee
In the case of nonzero $A_4(x) =  i \phi(x) \to 0$ instead of the singularities concentrated at a point in momentum space for any given $x$ we have singularities concentrated along the closed surfaces in momentum space. The form of these surfaces depends on $x$. More explicitly, we have spheres with the center at $p = A(x)$ and radius $|\phi_0(x)|$.
The Dirac operator becomes
\be
Q(\hat p-A(\hat x))=
\sum_{k} \Bigg[ i\gamma_k \xi_k +\frac{1}{2}\xi_k^2
\Bigg]
\ee
while
\be
\pd_{p_i}Q=\pd_{\xi_i}Q=
 i\gamma_i  +\xi_i
\ee
For the Green function we have
\be
\hat G(\^p-A(\^x))=
\frac{\left[\sum_{q} -i\gamma_q \xi_q +\frac{1}{2}\xi_q^2\right]}
{\sum_{k}\xi_k^2+\frac{1}{4}\(\sum_j \xi_j^2\)^2}
\ee
Topological invariant responsible for the CSE effect is given by
\begin{eqnarray}
&&\mathcal{N}=-\frac{1}{48 \pi^2 {\bf V}}
\int_{\Sigma_3}
\int d^3x
\tr \Bigg[\gamma^5
\Big[ G^{} \(d Q^{}\)   G^{}
\wedge  \(d Q^{}\)  G^{} \Big] \wedge d Q^{}
\Bigg]\label{NsimpleF}
\end{eqnarray}
Here surface $\Sigma_3$ surrounds all singularities of the Green functions at all values of $x$. The key point for the calculation of $\mathcal{N}$ is that we may deform the system smoothly removing the fields $A$, $\phi$ at all. This will bring us to a homogeneous system with $A=\phi=0$. Then we chose $\Sigma_3$ of the form of the $3$ sphere that surrounds point $p=0$. Then
\be
Q(\xi)=
\sum_{k} i\gamma_k \xi^k   + 2 \xi^2
\ee
and
\be
\pd_{p_i}Q=\pd_{\xi_i}Q=
i\gamma_i  +\xi\approx i\gamma_i
\ee
At the same time
\be
\hat G(\xi)=
\frac{-\sum_{q} i\gamma_q \xi^q + \xi^2/2 }
{\xi^2+\xi^4/4} \approx
\frac{-\sum_{q} i\gamma_q \xi^q}
{\xi^2}
\ee
and
\be
\pd_{p_i}G=\pd_{\xi_i}G=
-i\frac{\gamma_i - 2\xi^i  (\gamma \xi)/\xi^2}{\xi^2}
\ee
as a result integral over $x$ is irrelevant, and we have an integral over the surface of sphere ($d\sigma^i$ is a vector orthogonal to the surface of the sphere, its absolute value is equal to the area element):
\begin{eqnarray}
\mathcal{N}&=&\frac{1}{48 \pi^2}
\int_{\Sigma_3}
\tr \Big[\gamma^5
 G^{} \(d Q^{}\)  \wedge dG^{}
\wedge  \(d Q^{}\)\Big]\nonumber\\ & = &
\frac{\epsilon_{ijkl}}{48 \pi^2}
\int_{\Sigma_3} {d\sigma^i}
\tr \Big[\gamma^5
 G^{} \partial^j Q^{}  \partial^k G^{}
  \partial^l Q^{}\Big]\nonumber\\ & = &
\frac{\epsilon_{ijkl}}{48 \pi^2}
\int_{\Sigma_3} \frac{d\sigma^i}{\xi^4}
\tr \Big[\gamma^5
 \gamma^a \xi^a \gamma^j \gamma^k \gamma^l \Big]\nonumber\\ & = &
\frac{4 \epsilon_{ijkl}}{48 \pi^2}
\int_{\Sigma_3} \frac{d\sigma^i \xi^a}{\xi^4}
\epsilon^{ajkl}
\nonumber\\ & = &
\frac{24 \delta_{ia}}{48 \pi^2}
\int_{\Sigma_3} \frac{d\sigma^i \xi^a}{\xi^4}=1
\label{NsimpleF}
\end{eqnarray}

\end{document}